\documentclass[a4paper]{article}
\usepackage[a4paper,top=3cm,bottom=2cm,left=3cm,right=3cm,marginparwidth=1.75cm]{geometry}
\usepackage{amsfonts}

\usepackage{amssymb}
\usepackage{color}

\usepackage[all]{xy}
\usepackage{graphicx}
\usepackage{braket}
\usepackage{amsmath}
\usepackage{appendix}
\usepackage{graphicx}
\usepackage[colorinlistoftodos]{todonotes}
\usepackage{amsthm}
\usepackage{mathrsfs}
\usepackage{amssymb}
\usepackage{accents}
\usepackage{extarrows}
\usepackage{makecell}
\usepackage{bm}
\usepackage{authblk}
\usepackage{url}
\usepackage{hyperref}
\usepackage{cite}
\usepackage{eufrak}
\hypersetup{colorlinks,
            citecolor=black,
            linkcolor=black,
            urlcolor=green,
            pdftex}

\usepackage[english]{babel}
\usepackage[utf8x]{inputenc}
\usepackage[T1]{fontenc}

\title{Coherent intertwiner solution of simplicity constraint in all dimensional loop quantum gravity}
\author[1]{Gaoping Long \footnote{201731140005@mail.bnu.edu.cn}}
\author[1]{Chun-Yen Lin \footnote{cynlin@ucdavis.edu }}
\author[1]{Yongge Ma \footnote{mayg@bnu.edu.cn}\thanks{corresponding author}}

\affil[1]{Department of Physics, Beijing Normal University, Beijing 100875, China}
\date{}

\begin{document}

\maketitle

\begin{abstract}
  We propose a new treatment of the quantum simplicity constraints appearing in the general ${SO(D+1)}$ formulation of loop quantum gravity for the ${(1+D)}$-dimensional space-time. Instead of strongly imposing the constraints, we construct a specific form of weak solutions by employing the spin net-work states with specific ${SO(D+1)}$ coherent intertwiners. These states weakly satisfy the quantum simplicity constraint via the vanishing expectation values, and the quantum Gaussian constraints can be imposed strongly.  Remarkably, those specific ${SO(D+1)}$ coherent intertwiners used to construct our solutions have natural interpretations of the $D$-dimensional polytopes, commonly viewed as basic units of the discrete spatial geometry. Therefore, while the strong imposition of the quantum simplicity constraints leads to an over-constrained solution space, our weak solution space for the constraints may contain the correct semiclassical degrees of freedom for intrinsic geometry of the spatial hypersurfaces. Moreover, some concrete relations are established between our construction and other existing approaches in solving the simplicity constraints in all dimensional loop quantum gravity, providing valuable insights into this unresolved important issue.

%PACS numbers: 04.50.Kd, 04.20.Fy, 04.60.Pp.
\end{abstract}

\section{Introduction}

The theory of loop quantum gravity (LQG) in $(1+D)$-space-time dimensions can be constructed based on the Ashtekar formulation of canonical general relativity (GR). Such general formulation takes the same canonical variables as those of the $SO(D+1)$ Yang-Mills theory \cite{Bodendorfer:Ha}\cite{Bodendorfer:La} \cite{Bodendorfer:Qu}\cite{Bodendorfer:SgI}. In this formulation, canonical GR is governed by a first-class constraint system consisting of not only the familiar scalar and vector constraints, but also the additional Gaussian and the simplicity constraints. In the vacuum theory, the kinematical phase space is coordinatized by the conjugate pairs $(A_{aIJ},\pi^{bKL})$ of the spatial $SO(D+1)$ connection fields $A_{aIJ}$ and the densitized vector fields $\pi^{bKL}$. The Gaussian and simplicity constraints generate the gauge transformations of the canonical theory. Generating the local rotations introduced along with the internal components, the Gaussian constraints $\mathcal{G}^{IJ}$ play the expected role as in a standard Yang-Mills theory. However, it is crucial to notice that the internal gauge symmetry for the universal formulation corresponds to the $SO(D+1)$ group, rather than the $SO(D)$ rotation group for the $D$-dimensional tetrad variables. This is because the conjugate pair correspondence between the frame and connection variables happens naturally only for the privileged case of $D=3$, where the $SO(D)$ defining representation and the adjoin representation have the same dimension. To deal with this issue in the cases of higher dimensions, the universal connection formulation utilizes the Yang-Mills phase space of the augmented $SO(D+1)$, subject to the additional simplicity constraints $\mathcal{S}^{ab}_{IJKL}:=\pi^{a}_{[IJ}\pi^b_{KL]}$ generating the gauge transformations of the redundancy from the augmentation. Remarkably, The symplectic reduction of the $SO(D+1)$ Yang-Mills phase space with respect to Gaussian and simplicity constraints coincides with the familiar ADM phase space of GR in $(1+D)$-dimensional space-time. More specifically, after the simplectic reduction the variables $\pi^{bKL}$ give the spatial metric and $A_{aIJ}$ contains the degrees of freedom of the extrinsic curvature of the spatial hypersurface.

In the classical theory, the simplicity constraints are well understood. They can be rotated into one another under the $SO(D+1)$ gauge transformations. Thus they weakly commute with the Gaussian constraints on the simplicity constraint surface. Further, it is known that the scalar and vector constraints strongly commute with the simplicity constraints. Since the simplicity constraints at least weakly commute with all the rest of the constraints, it is instructive to study these special constraints separately.
In the case of $D=2$, the simplicity constraints are trivially satisfied. In the higher dimensional cases, the simplicity constraints can be properly imposed for the connection formulation to reproduce GR \cite{Bodendorfer:Ha}. However, in higher dimensional LQG, one is led to consider the quantized simplicity constraints acting on the kinematic states of quantum geometry. As it turns out, the algebra of the quantum simplicity constraints seems to be inevitably anomalous in a rather severe manner, such that it becomes questionable if the theory can recover the correct physical degrees of freedom. To gain new insights into these quantum simplicity constraints, in this paper we will explore a new approach of weakly imposing these quantum constraints.

Similar to the $(1+3)$-dimensional theory \cite{Ashtekar2012Background}\cite{Han2005FUNDAMENTAL}\cite{thiemann2007modern}\cite{rovelli2007quantum}, the quantum kinematics of LQG for arbitrary $(1+D)$ dimensions ($D\geq2$) is based on the elementary operators representing the classical $SO(D+1)$ holonomies of $A_{aIJ}$, and the fluxes of $\pi^{bKL}$. These operators generate the holonomy-flux quantum algebra, which is taken to be fundamental for the description of quantum geometry. The quantum Hilbert space supporting this operator algebra -- the completion of the space of cylindrical functions -- is spanned by a basis of states each of which is given by a network of hononomies, with a specific $SO(D+1)$ representation assigned to each edge of the network, and a specific coupling between the neighboring $SO(D+1)$ representations assigned to each vertex of the network. Under a given assignment of the representations to the edges, each successful assignment of $SO(D+1)$ invariant couplings to all of the vertices defines a spin network state. The spin network states, labeled by the $SO(D+1)$ representations and the $SO(D+1)$ invariant couplings called the intertwiners, form an orthonormal basis for the $SO(D+1)$ gauge invariant subspace in the space of cylindrical functions. Under the actions of well-defined geometric operators constructed from the flux operators, the labels of the spin network states give the quanta of spatial geometry, such as the (D-1)-areas. This provides the foundation of the quantum geometry of LQG. Thus, the standard loop quantization of any phase space function involves first representing the function with the holonomy and flux variables, and then naturally performing the quantization. The implementation of this loop representation leads to important physical implications resulted from the quantum geometric corrections. It also leads to new features and problems non-existing in the usual Fock representation of a Yang-Mills gauge theory.

For higher-dimensional LQG, how to implement the quantum simplicity constraints is an unresolved critical issue. The flux operators defined in the space of cylindrical functions become non-commuting, despite the fact that the classical flux variables that they represent obviously Poisson commute. This anomaly inevitably appears in the loop quantum representation of the simplicity constraints constructed with the flux operators. The result is that the classically commuting simplicity constraints do not even form a closed algebra after the loop quantization. More importantly, the transformations generated by these anomalous loop quantum simplicity constraints can happen between the states physically distinct in terms of the semiclassical limit. The strong imposition of the constraints thus leads to the over constrained physical states, which are unable to recover the full semiclassical degrees of freedom. The same problem also appears in $(1+3)$-dimensional spin-foam theory \cite{Freidel:BF}\cite{Barrett:1997Rsn}\cite{Engle:2007Fsv}\cite{Freidel:2007NSF}. This theory is thought of as a path-integral formulation of LQG, with each path as a history of transitions between the spin network states. The action involved is the Plebanski's action of GR expressed as a $(1+3)$-dimensional BF theory with certain additional constraints that are also called the simplicity constraints. Although these simplicity constraints are of different origin to the ones in our case, they have very similar structure and properties due to the shared function of reducing an augmented state space into a space carrying proper geometric meaning. The strong imposition of the quantum simplicity constraints in this context gives the Barrett-Crane model \cite{Barrett:1997Rsn} with the Barrett-Crane intertwiners (B-C intertwiners) known for its erroneous elimination of physical degrees of freedom. This problem has prompted people to find alternative treatments of simplicity constraints; for example, the classically equivalent linearized simplicity constraints \cite{Engle:2007LQGv} \cite{Engle:2007va}\cite{Livine:2007Nsfv}\cite{Kaminski:2009SFA} are put forward to replace the quadratic formulation of simplicity constraints that we mentioned above. More importantly, they are imposed weakly to deal with the anomaly. Although this way of dealing with anomaly is well known in the spin-foam theory, the analogous treatment of the quantum simplicity constraints in higher dimensional canonical LQG has yet to be thoroughly investigated.

In canonical LQG the various versions of quantum simplicity constraints have been discussed in many perspectives. Similar to the case of spin-foam theory, the strong imposition of the most natural quadratic quantum simplicity constraints will again lead to the B-C intertwiner \cite{Freidel:BF} which lacks the physical degrees of freedom. The alternative linearized quantum simplicity constraints have a strong solution space associated to each vertex, with the edges connecting to the vertex labelled with the so-called the simple representations \cite{Bodendorfer:2011onthe}. The vertex solutions for these linearized constraints are again equivalent to the B-C intertwiners.
%Also, the simple representations for each edge give only a one-dimensional space, lacking the physical degrees of freedom to account for the semiclassical geometry of the space.
Yet another treatment which has been studied is to choose a maximal subset of the simplicity constraint operators \cite{Bodendorfer:2011onthe} forming a closed and anomaly-free algebra. This subset can be interpreted as a specifying re-coupling scheme for the intertwiners at the vertices of a spin net-work state. Then, with assigning all the external and internal edges with simple representations, the gauge invariant intertwiners satisfying the chosen re-coupling scheme automatically solve this subset of the quantum simplicity constraints. These special intertwiners are called the ``simple" intertwiners.
%where the property ``simple" corresponds to the re-coupling scheme associated to the choice of the maximal closing subset.
The benefit of this treatment is that there is a unitary map between the $ Spin(4)$ simple intertwiner space and the $SU(2)$ intertwiner space, which is most commonly used for the privileged case with $D=3$. Also, this allows a comparison of the quantum simplicity constraints between the canonical theory and the four dimensional spin-foam theory \cite{Bodendorfer:2011onthe}. However, since such ``simple" intertwiners are based upon the specific form of the maximal closing subset of quantum vertex simplicity constraints, it remains unclear about the semiclassical meanings of the obtained intertwiner space and the unitary correspondence map.

These discussions motivate our approach in finding a weak solution space of the full quantum simplicity constraints for canonical LQG in arbitrary dimensions. We will review the most common forms of the quantum simplicity constraints and look for their weak solutions in the space of the cylindrical functions. Analogous to the treatment of the linear simplicity constraint in the $(1+3)$-dimensional spin-foam theory, our weak solutions will be constructed using a special from of the gauge-fixed intertwiners called the $SO(D+1)$ coherent intertwiners. This key construction is related to the idea that the intertwiner space solving all the ``internal" constraints should encode the shapes of a Euclidean polytope, representing the local structures of the discretized Reimannian geometry of space. This idea has been extensively studied in $(1+3)$-dimensional LQG in the $SU(2)$ formulation \cite{Bianchi:2010Polyhedra}, where specific $SU(2)$ coherent intertwiners has been shown to allow polytope interpretations, and the solution space of the $SU(2)$ quantum Gaussian constraints is indeed spanned by these specific coherent intertwiners. Here we achieve a conceptual and technical generalization of such framework to LQG under the universal $SO(D+1)$ formulation, where the additional quantum simplicity constraints appear. We will show that a subspace of the spin network states solving the quantum Gaussian constraints can also weakly solve the quantum simplicity constraints in the limit of large quantum number, via their expectation values and the minimal quantum uncertainties, when it is spanned by the states constructed with specific $SO(D+1)$ coherent intertwiners. Further, we will demonstrate that these specific $SO(D+1)$ coherent intertwiners are precisely the ones allowing the $D$-dimensional polytope interpretation as desired. The explicit form of these solutions will be given, and their properties relevant to the quantum anomalies will be discussed. Also, we will compare $Spin(4)$ simple coherent intertwiner in Bodendorfer-Thiemann's approach with the $SU(2)$ coherent intertwiner \cite{Dupuis:2010RSC} in Ashtekar-Lewandowski's approach of LQG, where the two different approaches might lead to different candidates of quantum GR in $(1+3)$-dimensions.

This paper is organized as below. In Section 2, we will review the standard quantum simplicity constraints, and also the required quantum kinematics obtained from the loop quantization on the Yang-Mills phase space of the universal connection formulation. We will consider both of the quadratic and linear versions of the quantum constraints. Also, we will point out the anomaly of the quantum vertex simplicity constraints and the necessity of a new strategy to deal with this problem. In Section 3, we will first introduce a toy model in quantum mechanics in order to understand properly the anomalous property of quantum vertex simplicity constraints and the motivation of our new strategy of using the $SO(D+1)$ coherent intertwiners. Then we will introduce the Perelomov $SO(D+1)$ coherent states and outline their important properties. The weak solutions of both the linear and quadratic quantum simplicity constraints will be constructed using this kind of $SO(D+1)$ coherent states, and their $SO(D+1)$ invariant counterparts solving further the quantum Gaussian constraints will be discussed. In Section 4, we will link our solution space to the space of shape of $D$-polytopes. This will be done in a concrete manner enabling the comparison between the construction in section 3 and the geometric quantization upon the space of shape of D-polytopes. In Section 5, our simple coherent intertwiners in $(1+3)$-dimensional Bodendorfer-Thiemann LQG will be compared with the coherent intertwiners in $(1+3)$-dimensional Ashtekar-Lewandowski LQG. It is shown that these two kinds of intertwiners have different properties, suggesting that these two approaches may lead to two distinct theories. Finally, our results will be summarized and some problems will be discussed briefly.

\section{Simplicity constraint in all dimensional canonical loop quantum gravity}

The connection formulation of $(1+D)$-dimensional GR is based on the Yang-Mills phase space, coordinatized by the $so(D+1)$-valued connection fields $A_{aIJ}$ and their conjugate momentum $\pi^{bKL}$ \cite{Bodendorfer:Ha}. Here we use the notation $a,b,...=1,2,...,D$ for the spatial tensorial indices and $I,J,... = 1,...,D+1$ for the $so(D+1)$ Lie algebra indices in the fundamental representation. As mentioned, in addition to the scalar and vector constraints, two sets of new constraints appear for this phase space. The $SO(D+1)$ Gaussian constraint appears as expected due to the $SO(D+1)$ gauge symmetry among the internal degrees of freedom. The simplicity constraint also arises from the adoption of the higher-dimensional rotation group, and they have the quadratic form given by \cite{Bodendorfer:Ha}
\begin{equation}
\pi^{a[IJ}\pi^{|b|KL]}\approx0.
\end{equation}
Classically, these conditions imply that each solution $\pi^{bKL}$ must take the bi-vector form as $\pi^{|b|KL}\approx n^{[K}E^{|b|L]}$, where $E^a_I$ is a hybrid vielbein density field, and $n^I$ is the corresponding (up to a sign) unit vector in the internal space, satisfying $n^In_I=1$, and $n^IE^a_I=0$. The spatial metric $q_{ab}$ is determined by the vielbein density field via $qq^{ab}=E^a_IE^{bI}$, with $q$ denoting the determinant of $q_{ab}$. The simplicity constraints can also be expressed in a linear form as \cite{Bodendorfer:SgI}
\begin{equation}
\mathcal{N}^{[I}\pi^{|a|JK]}\approx0,
\end{equation}
or equivalently
\begin{equation}
\bar{\pi}^{aIJ}:=\bar{\eta}^I_K\bar{\eta}^J_L\pi^{aKL}\approx0,\quad \text{with}\ \ \bar{\eta}^J_L:=\eta^J_L-\mathcal{N}^J\mathcal{N}_L,
\end{equation}
where an independent normalized internal vector field $\mathcal{N}^I$ is added as a phase space variable, together with its own canonical momentum $\mathcal{P}^J$.

Loop quantization of this phase space leads to the space of cylindrical functions as wave functions over the connection variables $A_{aIJ}$. An orthonormal basis for this space consists of elements labeled by: (1) a finite graph $\gamma$ in the spatial manifold consisting of a set of edges $\{e_\imath\}$ with their beginning and ending points connected at a set of vertices $\{v_n\}$; (2) a $SO(D+1)$ representation $\pi_{\Lambda_\imath}$ assigning to each of the edges; (3) an intertwiner $i_v$ assigning to each vertex $v$ connecting to the edges $\{e_{\imath_v}\}\subset \{e_{\imath}\}$. Each basis state $\Gamma_{\gamma,{\Lambda_\imath}, i_v}[A]$ as a wave function of the connection field is then given by
\begin{eqnarray}
\Gamma_{\gamma,{\Lambda_\imath}, i_v}[A]\equiv \bigotimes_{v}{i_v}\,\, \rhd\,\, \bigotimes_\imath h^{(\pi_{\Lambda_\imath})}_{e_\imath}[A],
\end{eqnarray}
where $h^{(\pi_{\Lambda_\imath})}_{e_\imath}$ denotes the holonomy of the edge $e_\imath$ in the representation $\pi_{\Lambda_\imath}$, and $\rhd$ denotes the contraction of the intertwiners with the holonomies. The wave function is then simply the product of the specified components of the holonomy matrices, selected by the projectors at the vertices. It has been shown that the Hilbert space of cylindrical functions contains all unconstrained wave functions of the connection fields \cite{Bodendorfer:Qu}\cite{Ashtekar2012Background}\cite{Han2005FUNDAMENTAL} \cite{thiemann2007modern}\cite{rovelli2007quantum}.

All the operators in this space can be constructed from the elementary set of holonomy and the flux operators. A holonomy operator $\hat{h}_e(A)$ and a flux operator $\hat{\pi}^{IJ}(S)$ act on a cylindrical function as
\begin{eqnarray}
\hat{h}_e(A)\cdot f_\gamma(A)&:=& h_e(A) f_\gamma(A),\\
\hat{\pi}^{IJ}(S)\cdot f_{\gamma_S}(A)&:=&-\mathbf{i}\hbar\kappa\beta\sum_{e\in E(\gamma_S)}\epsilon(e,S)R_e^{IJ}f_{\gamma_S}(A),
\end{eqnarray}
where $\gamma_S$ denotes a graph adapted to $S$ and equivalent to $\gamma$, $R_e^{IJ}:=\text{tr}((\tau^{IJ}h_e(0,1))^{\text{T}}\frac{\partial}{\partial h_e(0,1)})$ is the right invariant vector fields on $SO(D+1)$ associated to the edge $e$ of $\gamma_S$, with $\tau^{IJ}$ being an element of $so(D+1)$, and $\text{T}$ representing the transposition of the matrix, and $\epsilon(e,S)$ is defined by
\begin{equation}
\epsilon(e,S)=\left \{
\begin{aligned}
& +1, &\text{if } e\ \text{lies above the surface}\ S\ \text{and}\ b(e) \in S; \\
& -1, &\text{if } e\ \text{lies below the surface}\ S\ \text{and}\ b(e) \in S;\\
& 0,&\text{if } e\cap S=\varnothing\ \text{or}\ e \ \text{lies in} \ S;\ \ \ \ \ \ \ \ \ \ \ \ \
\end{aligned} \quad e\in E(\gamma_S),
\right.
\end{equation}
where $\gamma_S$ is such a graph that there are only outgoing edges at each true vertex, and $b(e)$ denotes the beginning point of the edge $e$.

As mentioned, in the classical theory there are two versions of the simplicity constraints --- the linear ones and the quadratic ones. Both forms can be loop quantized into the corresponding operators in the space of cylindrical functions. Here we will employ the most common scheme, which is to rewrite the set of simplicity constraints at each point $x$ of the spatial manifold in an infinitesimally smeared form using the flux variables. The set of smeared constraints for the point $x$ is given by replacing each factor of $\pi^{bKL} (x)$ appearing in the simplicity constraints by its flux $\hat{\pi}^{IJ}(S_x)$ over an arbitrary infinitesimal oriented surface $S_x$ containing $x$. Then, one may simply promote the flux variables into the flux operators and obtain the set of standard quantum simplicity constraint operators for any spatial point $x$. A strong solution state $f_{\gamma}$ for the quantum quadratic simplicity constraints at the point $x$ is supposed to satisfy
\begin{eqnarray}
\lim_{S_x, S'_x\rightarrow 0}\hat{\pi}^{[IJ}(S_x)\hat{\pi}^{KL]}(S'_x)f_{\gamma_{S_xS'_x}}(A)\approx0.
\end{eqnarray}
For a given graph $\gamma$, the actions of the flux operators become very simple, since they concern only the ways of intersections between the given edges of $\gamma$ and the surfaces $S_x$. Therefore, although there are infinitely many infinitesimal surfaces for every $x$ in space, the set of distinct actions by the quantum constraints on the specific cylindrical functions with a given graph reduces to only a finite set. Indeed, for each of the quantum constraint operators at a point $x$, there are only following three possibilities for its action: (1) One of the infinitesimal surfaces of the constraint operator does not intersect with the graph, and thus the states are annihilated by this constraint operator, and this includes all the cases when $x$ is not in the graph.  (2) The point $x$ lies in an edge of the graph, at which both of the flux surfaces of the constraint operator intersects with the edge. (3) The point $x$ coincides with a vertex of the graph, and the two flux surfaces may intersects with two different edges connecting to the vertex. Therefore, for the cylindrical functions associated with a certain graph, the original set of quantum simplicity constraints is equivalent to the new set of constraints imposed by the latter two cases of the actions, called the edge simplicity constraints and the vertex simplicity constraints. The edge simplicity constraints act as
\begin{eqnarray}
R_e^{[IJ}R_e^{KL]}f_{\gamma}(A)\approx0, \quad \forall e\in E(\gamma),
\end{eqnarray}
while the vertex simplicity constraints acts as
\begin{eqnarray}
\label{vertex constraint op}
R_e^{[IJ}R_{e'}^{KL]}f_{\gamma}(A)\approx0, \quad \forall e, e'\in E(\gamma)\ \text{and} \ b(e)=b(e')=v\in V(\gamma),
\end{eqnarray}
where $E(\gamma)$ and $V(\gamma)$ are the sets of edges and vertices of $\gamma$ respectively. It is known that a subspace of solutions to the edge simplicity constraints is spanned by the cylindrical functions with a specific subset $\{\pi_{N_\imath}\}\subset \{\pi_{\Lambda_\imath}\}$ of the irreducible representations \cite{Bodendorfer:Qu}, called the $SO(D+1)$ simple representations. The remaining task is thus to solve the vertex simplicity constraints in the subspace spanned by the basis \cite{Bodendorfer:Qu}\cite{Bodendorfer:2011onthe}
\begin{eqnarray}
\Gamma_{\gamma,{N_\imath}, p_v}[A]\equiv \bigotimes_{v}{p_v}\,\, \rhd\,\, \bigotimes_\imath h^{(\pi_{N_\imath})}_{e_\imath}[A],
\end{eqnarray}
where $N_\imath$ is a non-negative integer labelling a simple representation of $SO(D+1)$, and the assigned intertwiners coupling between the given simple representations are denoted by  $p_v\equiv\bigotimes_{\imath_v}\ket{N_{\imath_v}, \mathbf{M}_{\imath_v}}$, for their actions of selecting the $\mathbf{M}_{\imath_v}$ component of the holonomy $ h^{(\pi_{N_\imath})}_{e_\imath}$. As shown in \eqref{vertex constraint op}, the action of a vertex simplicity constraint is given by the composition of actions of the right invariant vector fields acting on the holonomies.
%For an arbitrary state $f_{\gamma}$ based on the graph $\gamma$ in the subspace of the simple representations, it has been shown that the complete building blocks of the vertex quadratic simplicity constraint operator acting on a vertex $v$ are given by \cite{Bodendorfer:Qu}
%\begin{equation}
%R^{e}_{[IJ}R^{e'}_{KL]}f_{\gamma}=0,\quad \forall e,e'\in\{e''\in E(\gamma); v=b(e'')\},
%\end{equation}
%where $f_{\gamma}$ is a cylindrical function defined on an adapted graph $\gamma$, and we have chosen the orientation of each edge to make sure that they are outgoing at $v$.
Since such actions involve inserting the corresponding $SO(D+1)$ generators into the vertex, two different quantum vertex simplicity constraints at the same vertex do not commute with each other in general. This is the anomaly of quantum quadratic simplicity constraints that mentioned above. It can be shown that the anomalous commutator between two vertex quadratic simplicity constraint operators is a linear combination of the following terms
\begin{equation}
\delta^{ABC\bar{E}}_{IJK\bar{M}}(R_{e''})_{AB}(R_{e})^{IJ}{(R_{e'})^K}_C,
\end{equation}
where $\bar{E}$ denotes $(D-3)$-tuple indexes and $\delta^{A_1A_2...A_n}_{I_1I_2...I_n}:=n!\delta^{[A_1}_{I_1}...\delta^{A_n]}_{I_n}$.

The linear simplicity constraints are quantized similarly with only one smearing surface. The resulted quantum linear simplicity constraint demands that \cite{Bodendorfer:2011onthe}

\begin{equation}\label{linear vertex constraint}
\hat{\mathcal{N}}^{[I}R_e^{JK]}f_\gamma\approx0,
\end{equation}
for all points of $
\gamma$. Here $\hat{\mathcal{N}}^{I}$ acts by multiplication and commutes with the right invariant vector fields. The above equation is equivalent to
\begin{equation}
\bar{R}_e^{IJ}f_\gamma\approx0, \quad \bar{R}_e^{IJ}:=\hat{\bar{\eta}}^I_K\hat{\bar{\eta}}^J_LR_e^{KL},\quad \hat{\bar{\eta}}^I_K:={\eta}^I_K-\hat{\mathcal{N}}^I\hat{\mathcal{N}}_K.
\end{equation}
It should be clear that the quantum constraint algebra in this case is given by the closed but non-commuting Lie algebra of the $SO(D+1)$ generators \cite{Bodendorfer:2011onthe}, and again  the algebra is anomalous due to the non-commutativity of insertions of the $SO(D+1)$ generators. Furthermore, although the commutator is small with higher power of $\hbar$, the trajectories of the kinematic wave packets generated by these quantum constraints deviate from the classical gauge orbits in a severe way, such that they can actually connect between different classical gauge orbits representing distinct physical states. This suggests that a strong imposition of the vertex linear simplicity constraints would over constrain the physical degrees of freedom. Besides, the quantum linear simplicity constraints act on edges problematically, and a group averaging is used to solve the problem. By this treatment, the linear constraints also enforce simple representations on the edges like the quadratic one \cite{Bodendorfer:2011onthe}. It is know that indeed the strong imposition of quantum quadratic simplicity constraint would give a one-dimensional solution space for the projector $p_v$, which is spanned by B-C intertwiners. It is also known that, the linear quantum simplicity has a one-dimensional strong solution space, and its group averaging is equivalent to the space spanned by B-C intertwiners. Therefore, the strong solution spaces for the quadratic and linear quantum simplicity constraints are identical, and such a space lacks the degrees of freedom that we expect for the recovery of the spatial geometry.

In the following sections, we will explore a different strategy of analyzing this problem, by using the $SO(D+1)$ coherent states for the vertex projectors to construct a meaningful type of weak solutions to the vertex simplicity constraints, fulfilling the following natural requirements.
\begin{itemize}
\item For these weak solutions, the expectation values of the quantum vertex simplicity constraints should be zero (or tend to zero in certain proper classical limit at least), and the expectation values of the (anomalous) constraint commutators should vanish similarly.

\item There should be minimal degeneracy for the orthogonal projection from the space of the weak solutions into the $SO(D+1)$ invariant intertwiner space. By this way, the weak solution space may be thought of as a gauge-fixed faithful representation of the $SO(D+1)$ invariant states, which not only weakly satisfy the simplicity constraints, but also solve the Gauss constraints.

\item The space of weak solutions should have proper degrees of freedom to describe semi-classical spatial geometry. More specifically, the $SO(D+1)$ coherent states solving the vertex constraints should be able to yield all the classical geometries of the $D$-polytopes dual to the vertex, and so the corresponding $SO(D+1)$ invariant coherent intertwiners could be labelled by the shapes of the dual polytopes.
\end{itemize}

\section{Coherent intertwiner solution of simplicity constraint}

Our weak solution treatment of the quantum simplicity constraints is motivated by the following simple example in quantum mechanics. Consider a particle moving in the unit $2$-dimensional sphere defined by $x^2+y^2+z^2=1$ embedded in a Euclidean space coordinatized by $(x,y,z)$. It is governed by a physical Hamiltonian $\tilde{\varepsilon}:=\frac{\vec{J}^2}{2}$ and a constraint $J_z=0$, where $\vec{J}$ is the angular momentum of the particle and $J_z$ is its component in $\frac{\partial}{\partial z}$ direction (we will call it the north direction). At the classical level, it is easy to see that all physically distinct trajectories of this particle under the condition $J_z=0$ are given by the various great circles passing through the north pole of the unit 2-dimensional sphere. Now suppose in the quantum theory we try to impose the constraint by constructing the physical Hilbert space of this particle as a solution space of the quantum constraint $\hat{J}_z\approx0$. In this case, the unconstrained Hilbert space is the homogeneous harmonic function space $\oplus_{j}V(j),j=0,1,2,...$, on ${}^2S$, where $V(j)$ is composed by homogeneous harmonic function with degree $j$. The strong solutions for the quantum constraint are easily given by the states $|j,0\rangle,j=0,1,2,...$ satisfying $\hat{J}_z|j,0\rangle=0$, and the weak solutions are provided by the coherent states $|j,\vec{n}_{\bar{z}}\rangle$ satisfying $\langle j,\vec{n}_{\bar{z}}|\hat{J}_z|j,\vec{n}_{\bar{z}}\rangle=0$ with minimal uncertainty, where $|\vec{n}_{\bar{z}}|=1,\vec{n}_{\bar{z}}\perp\frac{\partial}{\partial z},j=0,1,2,...$. Here the better choice in these two kinds of states is guided by the natural requirement that all classical states could be given by certain quantum states in some classical limit. It is easy to see that for a given $j\rightarrow\infty$, corresponding to the states with $|\vec{J}|=\sqrt{j(j+1)}\approx j$, the coherent states $|j,\vec{n}_{\bar{z}}\rangle$ could give the orbit corresponding to the classical state with angular momentum $\vec{J}=j\vec{n}_{\bar{z}}$, while the strong solution states $|j,0\rangle$ lack the degrees of freedom to account for the distinct classical orbits. Thus in terms of the degrees of freedom, the coherent states $|j,\vec{n}_{\bar{z}}\rangle$ are the better choice for a basis of the physical Hilbert space of this particle subjected to the constraint $J_z=0$. This example motivates us to construct a weak solution of quantum simplicity constraints based on coherent states.

\subsection{Coherent state in the simple representation space of SO(D+1)}

Using the familiar notations for the $SO(3)$ group, we denote the total angular momentum vector operator as $\vec{J}\equiv J_i \, \vec{n}^i$, where the set $\{\vec{n}^i; i=1,2,3\}$ denotes the orthonormal vector basis for the linear space of $so(3)$ and $J_i$ denotes the operator-valued coefficients, given by the corresponding $so(3)$ elements in a certain representation. In order to generalize this notation to the $SO(D+1)$ group, let us introduce a $so(D+1)$ basis of bi-vectors $\{\vec{n}^{\tilde{i}\tilde{j}}; \tilde{i},\tilde{j}=1,...,D+1\}$ consisting of the members given by ${n}_{IJ}^{\tilde{i}\tilde{j}}:=\delta^{[\tilde{i}}_I\delta^{\tilde{j}]}_J$, where $\{\delta_{\tilde{i}}^I=({\partial}/{\partial x^{\tilde{i}}})^I, \tilde{i}=1,...,D+1\}$ is just the orthogonal basis of the definition representation space of $SO(D+1)$. The total angular momentum operator in this basis can be written as  $\vec{X}=X_{\tilde{i}\tilde{j}}\vec{n}^{\tilde{i}\tilde{j}}$ with the $X_{\tilde{i}\tilde{j}}$ being the operator-valued coefficients given by the corresponding $so(D+1)$ elements in a certain representation. In the following we will use the component notation defined as
\begin{eqnarray}
X_{IJ}\equiv X_{\tilde{i}\tilde{j}}{n}_{IJ}^{\tilde{i}\tilde{j}}.
\end{eqnarray}
Note that in the case of defining representation we would have $X^{\tilde{i}\tilde{j}}$ given by $(X^{\tilde{i}\tilde{j}})^{\text{def}}_{KL}=2\delta^{[\tilde{i}}_K\delta^{\tilde{j}]}_L$. Under a $SO(D+1)$ transformation, $X^{\tilde{i}\tilde{j}}$ transforms as $gX^{\tilde{i}\tilde{j}}g^{-1}=2{g_I}^K\delta^{[\tilde{i}}_K\delta^{\tilde{j}]}_L{(g^{-1})^L}_J$ in the adjoint representation of $SO(D+1)$. Also, since $n_{IJ}^{\tilde{i}\tilde{j}}$ is a bi-vector in the  definition representation of $SO(D+1)$, and it is rotated by $g\in SO(D+1)$ as $g\circ n_{IJ}^{\tilde{i}\tilde{j}}:={g_I}^K\delta^{[\tilde{i}}_K\delta^{\tilde{j}]}_L{(g^{-1})^L}_J$. For the cylindrical functions with a fixed graph and the given representations on the edges, the space of the intertwiners at each vertex is naturally the product of the set of Hilbert spaces which support the operator $X_{IJ}$ in the given representation associated with one of the neighboring edges.

Recall that solving the edge simplicity constraints is to restrict to the space of cylindrical functions with only the simplicity representations on the edges. Therefore, to deal with the remaining vertex simplicity constraints in this space, we will restrict to the intertwiner space involving only the simple representations. Thus, we will study the projectors $p_v$ as the tensor products of the state vectors in the simple representation of $SO(D+1)$, and see how they behave under the actions of vertex simplicity constraint. As it is well known, each of the elements in the simple representation state space labelled by $N$ can be identified as a $SO(D+1)$ spherical harmonics function of degree $N$. To write down these spherical harmonics, we first construct an Cartesian coordinate system $(x_1,x_2,...,x_{D+1})$ in the defining representation space $\mathbb{R}^{D+1}$ of $SO(D+1)$. The operator $X_{\tilde{i}\tilde{j}}$ acts on a function in $\mathbb{R}^{D+1}$ simply as
\begin{equation}
X_{\tilde{i}\tilde{j}}\cdot f(x_1,x_2,...,x_{D+1})=(x_{\tilde{i}}\frac{\partial}{\partial x_{\tilde{j}}}-x_{\tilde{j}}\frac{\partial}{\partial x_{\tilde{i}}})f(x_1,x_2,...,x_{D+1}),
\end{equation}
where $\tilde{i},\tilde{j}=1,...,D+1$. The space $\mathfrak{H}^{N,D+1}$ of $SO(D+1)$ spherical harmonics of degree $N$, spanned by an orthonormal basis $\{Y^{\mathbf{M}}_N\}$ (where $\mathbf{M}=(M_1,...,M_{D-1})$, $N\geq M_1\geq,...,\geq |M_{D-1}|$), consists of the homogeneous polynomials in $\mathbb{R}^{D+1}$ of the same degree that satisfy the Laplace equations. Just as in the familiar $SU(2)$ case, the basis element with the maximal $\mathbf{M}$ is given by $(x_1+\textbf{i}x_2)^N$, which is an eigenvector of the operator $\vec{X}\cdot\vec{n}^{12}, (\vec{n}^{12}=n^{12}_{IJ}=\delta^{1}_{[I}\delta^{2}_{J]})$, with the eigenvalue of $\textbf{i}N$. We denote this state as $|N\mathbf{e}_1\rangle$, where $\mathbf{e}_k$ denotes the the generators in the dual space of the Cartan sub-algebra $\mathcal{C}$ of $SO(D+1)$, $\mathbf{e}_k(C_j)=\delta_{kj}$, here $\mathcal{C}$ is generated by $C_k=-\textbf{i}X_{2k-1, 2k}$ with $i,j,k=1,...,|\frac{D+1}{2}|$ \cite{Simon}. Also, we are given the following known properties of $|N\mathbf{e}_1\rangle$ when acted by $X_{\tilde{i}\tilde{j}}$:
\begin{equation}
X_{12}|N\mathbf{e}_1\rangle=\textbf{i}N|N\mathbf{e}_1\rangle,
\end{equation}
\begin{equation}
X_{\tilde{i}\tilde{j}}|N\mathbf{e}_1\rangle=0,\qquad \tilde{i},\tilde{j}\neq1,2,
\end{equation}
\begin{equation}
\langle N\mathbf{e}_1|X_{\tilde{i}\tilde{j}}|N\mathbf{e}_1\rangle=0,\qquad \tilde{i}=1,2,\tilde{j}\neq1,2,
\end{equation}
\begin{equation}
\langle N\mathbf{e}_1|X_{\tilde{i}\tilde{j}}X_{\tilde{i}\tilde{j}}|N\mathbf{e}_1\rangle =-\frac{N}{2},\qquad \tilde{i}=1,2,\tilde{j}\neq1,2,
\end{equation}
and
\begin{equation}
\Delta <X_{\tilde{i}\tilde{j}}> :=\sqrt{<X_{\tilde{i}\tilde{j}}X_{\tilde{i}\tilde{j}}>-(<X_{\tilde{i}\tilde{j}}>)^2}=\sqrt{-\frac{N}{2}},\qquad \tilde{i}=1,2,\tilde{j}\neq1,2,
\end{equation}
where we used the shorthand $<\alpha>\equiv\langle N\mathbf{e}_1|\alpha|N\mathbf{e}_1\rangle$. The above equations about the expectation values can be summarized as
\begin{equation}
<X_{IJ}>:=\langle N\mathbf{e}_1|X_{IJ}|N\mathbf{e}_1\rangle=2\textbf{i}N{n}^{12}_{IJ}.
\end{equation}
Further, the rest of the equations imply that the state $|N\mathbf{e}_1\rangle$ minimizes the uncertainty as
\begin{eqnarray}
\Delta(<X_{IJ}>)&:=&\sqrt{\sum_{I,J}\langle N\mathbf{e}_1|X_{IJ}X^{IJ}|N\mathbf{e}_1\rangle-\sum_{I,J}\langle N\mathbf{e}_1|X_{IJ}|N\mathbf{e}_1\rangle\langle N\mathbf{e}_1|X^{IJ}|N\mathbf{e}_1\rangle}\\\nonumber
&=&\sqrt{\sum_{\tilde{i},\tilde{j}}\langle N\mathbf{e}_1|X_{\tilde{i}\tilde{j}}X^{\tilde{i}\tilde{j}}|N\mathbf{e}_1\rangle-\sum_{\tilde{i},\tilde{j}}\langle N\mathbf{e}_1|X_{\tilde{i}\tilde{j}}|N\mathbf{e}_1\rangle\langle N\mathbf{e}_1|X^{\tilde{i}\tilde{j}}|N\mathbf{e}_1\rangle}\\\nonumber
&=&\sqrt{-2N(N+D-1)+2N^2}=\sqrt{-2N(D-1)},
\end{eqnarray}
with the relative uncertainty $\frac{|\Delta(<X_{IJ}>)|}{|<X_{IJ}>|}=\frac{\sqrt{D-1}}{\sqrt{N}}$ tends to $0$ when $N\rightarrow\infty$. It is in this sense, that these states are referred to as the $SO(D+1)$ coherent states. As in the usual case, the tensor product between two of such states gives the third one in the new representation of the combined $N$ numbers, i.e.,
\begin{equation}
|(N_1+N_2)\mathbf{e}_1\rangle=|N_1\mathbf{e}_1\rangle\otimes|N_2\mathbf{e}_1\rangle,
\end{equation}
which could be checked using the definition of $|N\mathbf{e}_1\rangle$. A general coherent state pointing in an arbitrary direction then follows from applying the $SO(D)_{x_1}$ transformations on $|N\mathbf{e}_1\rangle$ \cite{GeneralizedCoherentStates}, where $SO(D)_{x_1}$ is the maximal subgroup of $SO(D+1)$ stablizing the vector $\frac{\partial}{\partial x_1}$.

We have seen from the linear form of the vertex simplicity constraints, that a solution of the constraints should single out one privileged $SO(D)$ direction represented by the $\mathcal{N}^I$. To construct the weak solutions to the vertex constraints, we ultilize the coherent states related to $|N\mathbf{e}_1\rangle$ by elements $g_D\in SO(D)_{x_1}$. We denote these simple coherent states as $|N,X^{g_D}\rangle\equiv|N,g_D\rangle\equiv|N\mathbf{e}_1,g_D\rangle:=g_D|N\mathbf{e}_1\rangle$, where $X^{g_D}:=g_DX_{12}g^{-1}_D$. It follows directly from the above that
\begin{equation}
\label{inv vect expt 2}
<X_{IJ}>:=\langle N,g_D|X_{IJ}|N,g_D\rangle=2\textbf{i}Ng_Dn^{12}_{IJ}g^{-1}_D,
\end{equation}
and $|N,g_D\rangle$ minimize the uncertainty as
\begin{eqnarray}
&&\Delta(<X_{IJ}>)\\\nonumber
&:=&\sqrt{\langle N,g_D|X_{IJ}X^{IJ}|N,g_D\rangle-\langle N,g_D|X_{IJ}|N,g_D\rangle\langle N,g_D|X^{IJ}|N,g_D\rangle}\\\nonumber
&=&\sqrt{-2N(N+D-1)+2N^2}=\sqrt{-2N(D-1)},
\end{eqnarray}
with the relative size $\frac{|\Delta(<X_{IJ}>)|}{|<X_{IJ>}|}=\frac{\sqrt{D-1}}{\sqrt{N}}$ tending to $0$ when $N\rightarrow\infty$. The tensor products of these rotated coherent states also satisfy
\begin{equation}
|(N_1+N_2),g_D\rangle=|N_1,g_D\rangle\otimes|N_2,g_D\rangle.
\end{equation}
It is shown in next subsection that, by sharing one direction in their eigenvalues, these states weakly solve the vertex constraints, with the privileged direction given by this shared direction $\mathcal{N}^I\equiv (\frac{\partial}{\partial x_1})^I$.

\subsection{Simple coherent intertwiner}

Now we can construct the gauge-fixed simple coherent intertwiner for a $n_v$-valent vertex $v$. To define the operator $\hat{\mathcal{N}}^I$ appearing in the linear quantum constraints defined in \eqref{linear vertex constraint}, we also introduce a wave function for the quantum states of the sector of $\mathcal{N}$. Using the related coherent states introduced above, a set of solutions are immediately given in the form
\begin{equation}
\check{\mathcal{I}}_{s.c.}^{v,\delta}:=|\vec{N}_v,\vec{g}_D\rangle\cdot\delta^{S^D}(\mathcal{N}^I) :=\bigotimes_{\imath=1}^{n_v}|N_{\imath},g^{\imath}_D\rangle\cdot\delta^{S^D}(\mathcal{N}^I),
\end{equation}
where $\imath=1,...,n_v$, $\vec{N}_v:=(N_1,..,N_\imath,..,N_{n_v})$, $\vec{g}_D=(g_D^1,...,g_D^\imath,...,g_D^{n_v})$, $\mathcal{N}^I$ is a unit vector in $(D+1)$-dimensional Euclidean space which could be regarded as a point on $S^D$, and $\delta^{S^D}(\mathcal{N}^I)$ is the $\delta$ function on unit D-sphere $S^D$, which have the property that
\begin{equation}
\int _{S^D}\delta^{S^D}(p)f(p)dS^D=f(p_N),\quad p\in S^D,
\end{equation}
where $p_N$ is the north pole of $S^D$ corresponding to the unit vector ${(\frac{\partial}{\partial x_1})}^I$, and $f(p)$ is an arbitrary homogenous harmonic function on $S^D$. As requested in the natural requirements above, we want these solutions to give a faithful representation of the $SO(D+1)$ invariant states satisfying the quantum Gaussian constraints. To achieve this, we impose additional conditions among the above labels such that the coherent intertwiners give zero expectation values to all components of the $SO(D+1)$ generators. According to \eqref{inv vect expt 2} and the Liebniz rule of the generators acting on the individual edges at the vertex, this ``weak gauge-invariance" condition takes the form
\begin{eqnarray}
\sum_{\imath=1}^{n_v}N_\imath n^{IJ}( g^{\imath}_D)=0,
\end{eqnarray}
which is equivalent to
\begin{eqnarray}
\label{weak Gauss 2}
\sum_{\imath=1}^{n_v}N_\imath g^{\imath}_D\frac{\partial}{\partial x_2}=0,
\end{eqnarray}
where $n^{IJ}(g):=gn^{IJ}_{12}g^{-1}$. Here we denote by $\mathcal{\check{H}}^{s.c.\delta}_{\vec{N}_v}$ the space spanned by all of such $\mathcal{N}^I$-depended gauge-fixed simple coherent intertwiners $\check{\mathcal{I}}_{s.c.}^{v,\delta}$ with the given $(\vec{N}_v, \vec{g}_D)$ satisfying \eqref{weak Gauss 2}.

%The gauge invariant simple coherent intertwiners space can be obtain by performing the orthogonal projection on $\mathcal{\check{H}}^{s.c.\delta}_{\vec{N}_v}$ into the kernel of the quantum Gauss constraints, which is known to be equivalent to applying the $SO(D+1)$ group averaging on  $\mathcal{\check{H}}^{s.c.\delta}_{\vec{N}_v}$. 
The gauge invariant simple coherent intertwiners in the resulted space can be got by applying group averaging on the gauge-fixed ones $\check{\mathcal{I}}_{s.c.}^{v}:=|\vec{N}_v,\vec{g}_D\rangle :=\bigotimes_{\imath=1}^{n_v}|N_{\imath},g^{\imath}_D\rangle$ easily, which reads
\begin{equation}
\label{group ave}
\mathcal{I}^v_{s.c.}:=||\vec{N}_v,\vec{g}_D\rangle=\int_{SO(D+1)}dg \bigotimes_{\imath=1}^ng|N_{\imath},g^{\imath}_D\rangle=\int_{SO(D+1)}dg g^{\otimes n_v}|\vec{N}_v,\vec{g}_D\rangle.
\end{equation}
We will call this space spanned by the intertwiners $\mathcal{I}^v_{s.c.}$ the gauge invariant simple coherent intertwiner space $\mathcal{H}^{s.c.}_{\vec{N}_v}$. Note that the group averaging happens between the gauge-fixed projectors $\check{\mathcal{I}}_{s.c.}^{v}$ whose labels $\vec{g}_D$ are related by a left action of a $SO(D+1)$ element. Therefore we clearly see that $\mathcal{H}^{s.c.}_{\vec{N}_v} =\mathcal{\check{H}}^{s.c.}_{\vec{N}_v}/SO(D+1)^{\otimes n_v}$, where $\mathcal{\check{H}}^{s.c.}_{\vec{N}_v}$ is the gauge-fixed simple coherent intertwiner space spanned by $\check{\mathcal{I}}_{s.c.}^{v}$.
%The gauge fixed solution space faithful to the gauge invariant intertwiner solution space should be thus given by the representatives among $\check{\mathcal{I}}_{s.c.}^{v,\delta}$ with the representative labels among all the $\vec{g}_D$.

We note the following three important facts. First, the states in $\mathcal{\check{H}}^{s.c.\delta}_{\vec{N}_v}$ are weak solutions of the vertex linear quantum simplicity constraint $\hat{N}^{[I}R_{e_\imath}^{JK]}\approx0$. Indeed, for an arbitrary spin net-work state $|\gamma,\mathcal{\check{I}}_{v,\vec{N}_v,\vec{g}_D}^{s.c.\delta},...\rangle$ based on $\gamma\ni v$ and labelled with $\mathcal{\check{I}}_{v,\vec{N}_v,\vec{g}_D}^{s.c.\delta}$ at vertices $v$, we have
\begin{eqnarray}
\label{zero linear expt}
&&\langle\gamma,\mathcal{\check{I}}_{v,\vec{N}_v,\vec{g}_D}^{s.c.\delta},...|\hat{\mathcal{N}}^{[I}R_{e_\imath}^{JK]} |\gamma,\mathcal{\check{I}}_{v,\vec{N}_v,\vec{g}_D}^{s.c.\delta},...\rangle\\\nonumber &\simeq& \mathcal{N}^{[I}(p_N)\langle N_\imath,g^\imath_D|X^{JK]}|N_\imath,g^\imath_D\rangle=0,\quad \forall \imath,
\end{eqnarray}
where $\simeq$ means "proportion to", and we used the formula
\begin{equation}
\int_{S^D}\mathcal{N}^I\delta^{S^D}(\mathcal{N}^I)d\nu(\mathcal{N}^I)\simeq \mathcal{N}^{I}(p_N),
\end{equation}
where $d\nu(\mathcal{N}^I)=dS^D$ is the invariant measure on $S^D$. Second, the states in $\mathcal{\check{H}}^{s.c.}_{\vec{N}_v}$ are weak solutions of the vertex quadratic quantum simplicity constraints, which are formed with the building blocks $R_{e_\imath}^{[IJ}R_{e_\jmath}^{KL]}$. This can be easily checked by evaluating
\begin{eqnarray}
\label{zero quadratic expt}
&&\langle\gamma,\mathcal{\check{I}}_{v,\vec{N}_v,\vec{g}_D}^{s.c.},...|R_{e_\imath}^{[IJ}R_{e_\jmath}^{KL]} |\gamma,\mathcal{\check{I}}_{v,\vec{N}_v,\vec{g}_D}^{s.c.},...\rangle\\\nonumber &\simeq&\langle N_\imath,g^\imath_D|X^{[IJ}|N_\imath,g^\imath_D\rangle\langle N_\jmath,g^\jmath_D|X^{KL]}|N_\jmath,g^\jmath_D\rangle\\\nonumber
&=&0,\quad \forall \imath,\jmath.
\end{eqnarray}
Third, in the large $N_\imath$ limit the gauge invariant $\mathcal{H}^{s.c.}_{\vec{N}_v}$ tends to weak solutions of the vertex quadratic quantum simplicity constraints, i.e.,
\begin{equation}
\lim_{N_\imath\to \infty}\langle\gamma,\mathcal{I}_{v,\vec{N}_v,\vec{g}_D}^{s.c.},...|R_{e_{\jmath_1}}^{[IJ}R_{e_{\jmath_2}}^{KL]} |\gamma,\mathcal{I}_{v,\vec{N}_v,\vec{g}_D}^{s.c.},...\rangle=0.
\end{equation}

It is important to note that the key factors $\langle N_\jmath,g^\jmath_D|X^{KL}|N_\jmath,g^\jmath_D\rangle$ in Eqs. \eqref{zero linear expt} and \eqref{zero quadratic expt} are determined by the properties of the coherent state $|N_\jmath,g^\jmath_D\rangle$, and thus the quantum constraints are satisfied up to the fluctuations of minimal uncertainty. Consequentially, one can check that the anomalous term $\delta^{ABC\bar{E}}_{IJK\bar{M}}(R_{e''})_{AB}(R_{e})^{IJ}{(R_{e'})^K}_C$ also has zero expectation value for the coherent intertwiner and its uncertainty is again minimal as coming from the key factors $\langle N_\jmath,g^\jmath_D|X^{KL}|N_\jmath,g^\jmath_D\rangle$. It should be also noted that the gauge invariant simple coherent intertwiners $\mathcal{I}_{v,\vec{N}_v,\vec{g}_D}^{s.c.}$ weakly solve the constraints only in the large $N_\imath$ limit. This can be checked by evaluating the expectation values of the key factor $R_{e_{\jmath_1}}^{[IJ}R_{e_{\jmath_2}}^{KL]}$ given by them as
\begin{eqnarray}\label{gauge inv expect 0}
&&\langle\gamma,\mathcal{I}_{v,\vec{N}_v,\vec{g}_D}^{s.c.},...|R_{e_{\jmath_1}}^{[IJ}R_{e_{\jmath_2}}^{KL]} |\gamma,\mathcal{I}_{v,\vec{N}_v,\vec{g}_D}^{s.c.},...\rangle\\\nonumber &\simeq&\int_{SO(D+1)}\int_{SO(D+1)} \prod_{\imath\neq\jmath_1,\jmath_2,\imath=1}^{n_v} \langle N_{\imath},g^{\imath}_D|h^{-1}g|N_{\imath},g^{\imath}_D\rangle\\\nonumber
&&\cdot \langle N_{\jmath_1},g^{\jmath_1}_D|h^{-1}X^{[IJ}g|N_{\jmath_1},g^{\jmath_1}_D\rangle\cdot \langle N_{\jmath_2},g^{\jmath_2}_D|h^{-1} X^{KL]}g|N_{\jmath_2},g^{\jmath_2}_D\rangle dgdh \\\nonumber&=&\int_{SO(D+1)}\int_{SO(D+1)} \prod_{\imath\neq\jmath_1,\jmath_2,\imath=1}^{n_v} \langle N_{\imath},g^{\imath}_D|g|N_{\imath},g^{\imath}_D\rangle\\\nonumber
&&\cdot \langle N_{\jmath_1},g^{\jmath_1}_D|h^{-1}X^{[IJ}hg|N_{\jmath_1},g^{\jmath_1}_D\rangle\cdot \langle N_{\jmath_2},g^{\jmath_2}_D|h^{-1} X^{KL]}hg|N_{\jmath_2},g^{\jmath_2}_D\rangle dgdh,
\end{eqnarray}

where we used that
\begin{eqnarray}
R_{e_{\jmath_1}}^{[IJ}R_{e_{\jmath_2}}^{KL]} \mathcal{I}_{v,\vec{N}_v,\vec{g}_D}^{s.c.} &=&\int_{SO(D+1)}\bigotimes_{\imath\neq\jmath_1,\jmath_2,\imath=1}^{n_v} g|N_{\imath},g^{\imath}_D\rangle\\\nonumber
&&\otimes \tau^{[IJ}g|N_{\jmath_1},g^{\jmath_1}_D\rangle\otimes \tau^{KL]}g|N_{\jmath_2},g^{\jmath_2}_D\rangle dg.
\end{eqnarray}
It is shown in the Appendix that Eq. \eqref{gauge inv expect 0} tends to zero only in the large $N_\imath$ limit.

\subsection{Relationship between weak and strong solutions of linear simplicity constraints}

 In the quantum mechanical example given at the beginning of this section, the relationship between the strong and weak solutions of $\hat{J}_z\approx0$ is given by
\begin{equation}\label{j0integral}
|j,0\rangle=\frac{1}{c_j}\int_{SO(2)_z}dg g|j, \vec{n}_{\bar{z}}\rangle,\quad g\in SO(2)_z,\quad \forall \vec{n}_{\bar{z}}
\end{equation}
where $c_j=\langle j,0|j,\vec{n}_{\bar{z}}\rangle$, $SO(2)_z$ is the subgroup of $SO(3)$ which fixes the vector $\frac{\partial}{\partial z}$, and the measure $dg$ satisfies $\int_{SO(2)_z}dg=1$. The integration in Eq.\eqref{j0integral} can be regarded as the average taken over the transformations generated by the quantum constraint $\hat{J}_z$. The validity of Eq.\eqref{j0integral} can be checked easily by inserting the completely orthonormal basis $\{\langle j,m||m=0,\pm1,...,\pm j\}$ in the dual space of $V(j)$ on both sides of the equation. Similar to 3-dimensional case, one can check the following identity for $SO(D+1)$ group,
\begin{equation}
|N,\mathbf{0}\rangle=\frac{1}{c_N}\int_{SO(D)_{x_1}}dgg|N,g_D\rangle,\quad g\in SO(D)_{x_1},\quad c_N=\langle N,\mathbf{0}|N,g_D\rangle, \quad \forall g_D\in SO(D)_{x_1},
\end{equation}
by expanding both sides of the equation with the complete orthonormal basis $
\{\langle N,\mathbf{M}||\mathbf{M}=(M_1,M_2,...,M_{D-1}),N\geq M\geq M_1\geq M_2\geq...\geq |M_{D-1}|\}$ in the dual space of $\mathfrak{H}^{N,D+1}$, where the measure $dg$ satisfies $\int_{SO(D)_{x_1}}dg=1$. These states indeed provide the strong solutions to the linear vertex simplicity constraints in the form
\begin{equation}
\mathcal{I}^{\text{B.T.}}_v(\vec{N}):=\bigotimes_{\imath=1}^{n_v} |N_\imath,\mathbf{0}\rangle\cdot\delta^{S^D}(\mathcal{N}^I),\quad \vec{N}:=(N_1,N_2,...,N_{n_v}),
\end{equation}
which is equivalent to the strong solution of linear quantum simplicity constraint given by Bodendofer and Thiemann \cite{Bodendorfer:2011onthe}.
Then we have
\begin{eqnarray}\label{BT}
\mathcal{I}^{\text{B.T.}}_v(\vec{N})&=&\frac{1}{c_{N_1}\cdot...\cdot c_{N_{n_v}}}\bigotimes_{\imath=1}^{n_v}(\int_{SO(D)_{x_1}}dg_\imath g_{\imath}|N_\imath,g_D^\imath\rangle)\cdot\delta^{S^D}(\mathcal{N}^I)\\\nonumber
&\equiv&\frac{1}{c_{N_1}\cdot...\cdot c_{N_{n_v}}}\bigotimes_{\imath=1}^{n_v}(\int_{SO(D)_{x_1}}dg_\imath g_{\imath})\circ\bigotimes_{\imath=1}^{n_v}|N_{\imath},g^{\imath}_D\rangle\cdot\delta^{S^D}(\mathcal{N}^I)\\\nonumber
&=&\frac{1}{c_{N_1}\cdot...\cdot c_{N_{n_v}}}\bigotimes_{\imath=1}^{n_v}(\int_{SO(D)_{x_1}}dg_\imath g_{\imath})\circ\check{\mathcal{I}}_{s.c.}^{v,\delta},
\end{eqnarray}
where the integrals in right side of the equation are the averages of the transformations induced by the quantum linear simplicity constraint $\bar{R}^{IJ}_{e_\imath}:=\hat{\eta}^I_K\hat{\eta}^J_LR^{KL}_{e_\imath}\approx0$, $\hat{\eta}^I_K:=\eta^I_K-\hat{\mathcal{N}}^I\hat{\mathcal{N}}_K$. Eq.\eqref{BT} implies that the weak solutions of the vertex simplicity constraint that we introduced can be regarded as ``gauge" (of quantum simplicity constraint) fixed formulation of the strong solutions, and here the ``gauge" degrees of freedom have true physical meanings.
%$\langle N_{\jmath_1},\underline{g}^{\jmath_1}_D|\otimes\langle N_{\jmath_2},\underline{g}^{\jmath_2}_D|$ is the projection of $\bigotimes_{\imath\neq\jmath_1,\jmath_2,\imath=1}^{n_v} \langle N_{\imath},g^{\imath}_D|$ into the space $\mathfrak{H}^{D+1,N_{\jmath_1}}\otimes \mathfrak{H}^{D+1,N_{\jmath_1}}$, and
%and
%\begin{eqnarray}
%&&\int_{SO(D+1)}dg\langle N,g^{\imath_1}_D|g|N,g^{\imath_2}_D\rangle\langle N',g^{\imath_3}_D|g|N',g^{\imath_4}_D\rangle\\\nonumber
%&=&\frac{1}{dim.(\mathfrak{H}^{D+1,N})}\langle N,g^{\imath_1}_D|N',g^{\imath_3}_D\rangle\langle N,g^{\imath_2}_D|N',g^{\imath_4}_D\rangle.
%\end{eqnarray}
%neglct
%\begin{eqnarray}
%&<&\int_{SO(D+1)}(\langle N_{\jmath_1},\underline{g}^{\jmath_1}_D|\otimes\langle N_{\jmath_2},\underline{g}^{\jmath_2}_D|)(|N_{\jmath_1},g^{\jmath_1}_D\rangle \otimes|N_{\jmath_2},g^{\jmath_2}_D\rangle)\\\nonumber
%&&\cdot(\langle N_{\jmath_1},\underline{g}^{\jmath_1}_D|\otimes\langle N_{\jmath_2},\underline{g}^{\jmath_2}_D|)(h^{-1}X^{[IJ}h|N_{\jmath_1},g^{\jmath_1}_D\rangle \otimes h^{-1}X^{KL]}h|N_{\jmath_2},g^{\jmath_2}_D\rangle)dh
%\end{eqnarray}

\section{Polytopes in D-dimensional Euclidean space and simple coherent intertwiner}

%The phase space of a free particle which move on a unit D-sphere $S^D$ is given by $\mathfrak{M}=T^{\ast}S^D-\{$0-section$\}:= \{(p,q)\in T^{\ast}\textbf{R}^{n+1}||x|=1,x\cdot y=0\}$, where $\textbf{R}^{n+1}$ and $T^{\ast}\textbf{R}^{n+1}$ are the $(n + 1)$-space and its cotangent bundle with coordinates $p= (p_1,... ,p_{n+1})$ and $(p, q) =(p_1,... ,p_{n+1},q_1,... ,q_{n+1})$ respectively.
In a quantum geometric interpretation of spin network states in the $SU(2)$ formulation of $(1+3)$-dimensional LQG, the quantum geometry of the 3-dimensional space is given by a set of quantum 3-polytopes glued together in a consistent manner. Specifically, each of the 3-polytopes is dual to a vertex, with its faces dual to the edges adjacent to the vertex. The states of the intertwiners determine the geometries of the 3-polytopes, while the edge degrees of freedom specify the gluing of the polytopes. The key fact allowing this interpretation is that the intertwiners satisfying the quantum Gaussian constraints capture precisely the degrees of freedom of the shapes of the corresponding polytopes. In this section we are going to show that, in the universal formulation with the quantum simplicity constraints, there is again an analogous correspondence between the the $D$-polytopes and our gauge invariant simple coherent intertwiners in the weak solution space $\mathcal{{H}}^{s.c.}_{\vec{N}_v}$ of the quantum simplicity constraints. Our conclusion is that the simple coherent intertwiner space may be thought of as the \textbf{quantum counterpart} of the space of shapes of $D$-polytopes in $(D+1)$-dimensional Euclidean space, with given areas of each $(D-1)$-dimensional faces ($(D-1)$-faces). Here we use the terminology ``quantum counterpart'' rather than ``quantum Hilbert space'' because we have not directly quantized the space of shapes of polytopes to obtain the simple coherent intertwiners space. The precise meaning of ``quantum counterpart'' in this context will be explained in the end of this section.

\subsection{Classical polytopes in D-dimensional Euclidean space}

The starting point of our analysis is the well-known fact that, given a set of normalized basis vectors $(V^I_1,...,V^I_n)$ in the $D$-dimensional Euclidean space and a set of positive numbers $(A_1,...,A_n)$ satisfying the closure condition $\sum_{\imath=1}^nA_\imath V^I_\imath=0$, there is a unique D-polytope with $n$ (D-1)-faces having the areas $(A_1,...,A_n)$ and normal vectors $(V^I_1,...,V^I_n)$. This result is guaranteed by the following two theorems \cite{Minkowski}:

(i) \textbf{Theorem} (H. Minkowski uniqueness theorem). Let $D\geq2$ and let two convex polytopes in $\mathbb{R}^D$ be such that, for every (D-1)-face of each of the polytopes, the parallel face of the other polytope has the same (D-1)-dimensional volume. Then the polytopes are congruent and parallel to each other.

(ii) \textbf{Theorem} (H. Minkowski existence theorem). Let $D\geq2$ and let $V^I_1,...,V^I_n$ be unit vectors in $\mathbb{R}^D$ which do not lie in a closed half-space bounded by a hyperplane passing through the origin. Let $A_1,...,A_n$ be positive real numbers such that $\sum_{\imath=1}^nA_\imath V^I_\imath=0$. Denote by $\vec{A}$ and $\vec{V}^I$ the sets $\{A_1,...,A_n\}$ and $\{V^I_1,...,V^I_n\}$ respectively. Then there exists a convex polytope $p(\vec{A},\vec{V}^I)$ in $\mathbb{R}^D$ such that the vectors $V^I_1,...,V^I_n$ (and only they) are the unit outward normal vectors to the (D-1)-faces of $p(\vec{A},\vec{V}^I)$ and the (D-1)-dimensional volumes of the corresponding faces are equal to $A_1,...,A_n$.

In the $SU(2)$ formulation of the (1+3)-dimensional theory, the closure condition manifests through the Gaussian constraint, with the normal vectors in the 3-dimensional space identified with the $su(2)$ Lie algebra elements that satisfy the constraints. However, in dealing with the higher dimensional cases, we have to use the $SO(D+1)$ group whose Lie algebra is no longer identifiable to the D-dimensional vector space. So let us first tailor our application of the D-dimensional Minkowski theorem to the $SO(D+1)$ context. Consider a set of normalized $so(D+1)$ elements $\vec{V}^{IJ}:=\{V_{1}^{IJ},...,V_{n}^{IJ}\}$ which span a D-dimensional subspace of the algebra, and suppose that they also satisfy the conditions $V_\imath^{[IJ}V_\jmath^{KL]}=0$ and $\sum_\imath A_\imath V_{\imath}^{IJ}=0$. As mentioned in section 2, the first condition implies that the elements take the form $V_{\imath}^{IJ}:=\sqrt{2}\mathcal{N}^{[I}V^{J]}_\imath$, corresponding to a set of normalized vectors $\{\mathcal{N}^I,V^I_1,...,V^I_n\}$ in the $(D+1)$-dimensional definition vector space of $SO(D+1)$ with $\mathcal{N}_IV^I_\imath=0,\quad \forall \imath=1,...,n$. It is clear that the vectors $\{V^I_1,...,V^I_n\}$ span a D-dimensional vector space which is the subspace of the $(D+1)$-dimensional space orthogonal to $\mathcal{N}_I$. The second condition further implies $\sum_\imath A_\imath V_{\imath}^{J}=0$, which then allows us to assign a polytope to the set $\vec{V}^{IJ}$ as given by $p(\vec{A},\vec{V}^{IJ})\equiv p(\vec{A},\vec{V}^{I})$. Conversely, it is straightforward to show that we can embed each $D$-polytope into the $(D+1)$-dimensional vector space, so that the normals of their faces are orthogonal to an additional dimension given by $\mathcal{N}_I$, and thus the embedded polytope takes the form $p(\vec{A},\vec{V}^{I})$. The set $\mathcal{N}_I$ and $V^{I}$ then specify a set $V_{\imath}^{IJ}:=\sqrt{2}\mathcal{N}^{[I}V^{J]}_\imath$ satisfying the above conditions. This establishes the correspondence between the space of the $D$-polytopes and the sets of the $so(D+1)$ basis elements satisfying the simplicity constraints and closure condition (or Gaussian constraint). Such correspondence has been also introduced in \cite{Bodendorfer:2013sja}.

Referring to the above setting, we now construct the phase space of one single elementary $(D-1)$-dimensional face in the $(D+1)$-dimensional space, before imposing any of the constraints. This face labeled by $\imath$ is characterized by its two normals $\mathcal{N}^I_\imath$ and $V^{I}_\imath$ and its area $A_\imath$. The normals correspond to the $so(D+1)$ elements $V_{\imath}^{IJ}$ above that transform in the adjoint representation of $SO(D+1)$. Since there is a subgroup $SO(D-1)$ of $SO(D+1)$ preserving both of the normals and another subgroup $SO(2)$ preserving the space spanned by the two normals, we conclude that a proper coordinate space of the $(D-1)$-faces is actually given by the pairs $(Q^\imath_{D-1}, A_\imath)$ with $Q^\imath_{D-1}$ being the quotient manifold
\begin{equation}\label{QD-1}
Q^\imath_{D-1}:=SO(D+1)/(SO(D-1)\times SO(2)).
\end{equation}
With a set of specified values for $A_\imath$, we may assign to the corresponding subspace $(Q^\imath_{D-1}, A_\imath)|_{(\text{fixed} A_\imath)}$ a symplectic form by using $\Omega_{A_\imath^2/2}:=A_\imath\Omega$, where $\Omega$ is the natural Kahler form on the compact Kahler manifold $Q^\imath_{D-1}$. The phase space of $n$ faces with the given areas $(A_1,...,A_n)$ is simply $n$ copies of the phase space above. The space of shapes of D-polytopes with the given areas $(A_1,...,A_n)$ can then be viewed as a constraint surface in the phase space, module the overall rotation of $SO(D+1)$ preserving the shape. The resulted space is given by
\begin{eqnarray}
&&\mathfrak{P}^{s.}_{\vec{A}}:=\{(A_1V^{IJ}_1,A_2V^{IJ}_2,...,A_nV^{IJ}_n)\in Q_{D-1}(A_1)\times Q_{D-1}(A_2) \times\\\nonumber&&...\times Q_{D-1}(A_n)
| \sum_{\imath=1}^nA_\imath V^{IJ}_\imath=0,\quad V_\imath^{[IJ}V_\jmath^{KL]}=0\}/SO(D+1),
\end{eqnarray}
where $V^{IJ}_\imath$ are rotated by $SO(D+1)$ with adjoint representation, and $\mathfrak{P}^{s.}_{\vec{A}}$ has the natural form induced by $\Omega_{A^2_1/2}\times...\times \Omega_{A^2_n/2}$.

\subsection{Geometric quantization of polytopes in D-dimensional Euclidean space }

The above prescription of the polytopes as the points in the constraint surface of the phase space $(Q_{D-1}(A_1)\times Q_{D-1}(A_2) \times...\times Q_{D-1}(A_n))$ suggests a way to quantize the polytopes. One may first quantize the phase space into a unconstrained Hilbert space describing the sets of ``quantum faces", and then try to find the quantum polytope states satisfying the quantum Gaussian and simplicity constraints defined in the unconstrained Hilbert space. We observe that our phase space of a single face with a given area can be identified with the phase space of angular momentum $L_{IJ}$ of a particle moved on unit D-dimensional sphere with a given energy, with the $\mathcal{N}^I$ corresponding to the location of the particle on the sphere, the $A_\imath V_\imath^J$ corresponding to the velocity, and the area $A^2_\imath/2$ corresponding to the energy $\varepsilon:=\frac{\tilde{\Delta}}{2}:=\frac{1/2L_{IJ}L^{IJ}}{2}$. This phase space is a compact Kahler manifold $Q_{D-1}$ defined by \eqref{QD-1} with the Kahler form $\Omega_\varepsilon:=\sqrt{2\varepsilon}\Omega$. The topology of this phase space is not a co-tangent bundle and there is no canonical coordinate on this phase space. Hence the usual canonical quantization approach cannot be applied. Nevertheless, a complete quantization of this phase space has been achieved using the the approach of geometric quantization, under the quantum condition on the energy $\varepsilon$ \cite{Geometricquantization}\cite{Geometricquantization2}:
\begin{equation}\label{N+D-1/2}
\sqrt{2\varepsilon}=N+\frac{D-1}{2}, \quad N=0,1,2,...
\end{equation}
for the existence of the corresponding Hilbert space. The dimension of this Hilbert space is given by:
\begin{equation}
\textrm{dim}(\mathcal{H}^N(Q_{D-1},\Omega_\varepsilon)) =\frac{(2N+D-1)(D+N-2)!}{(D-1)!N!}.
\end{equation}
This dimensionality is as same as that of the $SO(D+1)$ simple representation space $\mathfrak{H}^{N,D+1}$ with the Casimir value of $N(N+D-1)$. However, one must note from Eq.\eqref{N+D-1/2}, that the norm of angular momentum $|\tilde{\Delta}|$ corresponding to quantum number $N$ is given by $2\varepsilon=(N+(D-1)/2)^2$, which differs from the value $N(N+D-1)$ given by the $SO(D+1)$ Casimir operator $\Delta:=-1/2X_{IJ}X^{IJ}$ appeared in the canonical quantization in LQG. This means that, although the geometric quantization and the loop quantization agree in the degrees of freedom of the $SO(D+1)$ angular momentum, the spectra for the norm of the angular momentum in the two approaches differ by a quantum correction. Here we will neglect this quantum correction in $\hat{\varepsilon}:=\frac{\Delta}{2}+\frac{(D-1)^2}{8}$ for simplicity and choose the eigenvalue of $SO(D+1)$ Casimir operator $\Delta$ as the spectrum of the energy operator such that $\hat{\varepsilon}=\frac{\Delta}{2}$. Up to this ignored correction to the spectrum, we can now identify the quantum space $\mathcal{H}^N(Q_{D-1},\Omega_\varepsilon)$ from the geometric quantization with the quantum space $\mathfrak{H}^{D+1,N}$ from loop quantization.

Let us look into the significance of this identification. In LQG one performs a canonical quantization to the Yang-Mills phase space $(\pi, A)$ written in the loop variables and obtains the space of cylindrical functions. In this space one then applies a part of the quantum simplicity constraints, the edge simplicity constraints, and finds the solution space to be just the subspace spanned by the cylindrical functions with edges assigned with the simple representations. Particularly, the space of the vertex projectors for this subspace is given by the space $\times_{\imath=1}^n\mathfrak{H}^{D+1,N_\imath}$. In the above, it is shown that this space turns out to be exactly the one obtained by the geometric quantization on the phase space of the elementary faces. It is desirable if the exact correspondence would be maximally preserved when the further constraints are imposed on both sides, since this would correspond the quantum space of the polytopes from the geometric quantization to our proposed space of solutions. To show that this expectation can be indeed realized by two stages, we will first impose the closure constraints and then the simplicity constraints starting from the phase space of the faces.

%The phase space $\mathfrak{Q}^k_{D-1}$ of the angular momentum $L_{IJ},\quad |L_{IJ}|=k$ of this particle could be given by $(\mathfrak{Q},k\Omega)$, where $\mathfrak{Q}_{D-1}:=SO(D+1)/(SO(D-1)\times SO(2))$. $\mathfrak{M}$ can be quantized as $L^2(S^D)$ following the geometric quantization procedure, and the angular momentum $L_{IJ}$ will be quantized as operators $i X_{IJ}$ which act on elements of $L^2(S^D)$ as Lie-algebra element of $so(D+1)$. In detail, the the phase space $\mathfrak{Q}(N)$ of the angular momentum $L_{IJ}, |L_{IJ}|=N$ will be quantized as $\mathfrak{H}^{D+1,N}$ and $|L_{IJ}|=|X_{IJ}|=\sqrt{-1/2X_{IJ}X^{IJ}}$ will be modified as $\sqrt{N(N+D-1)}, N=1,2,3,...$.

In the first stage, the correspondence remains exact. The key fact here is that the closure constraints are also the generators of the $SO(D+1)$ gauge symmetry representing the overall rotation of a group of faces. Thus the space of the gauge orbits on the constraint surface is simply the reduced phase space obtained by the usual symplectic reduction. More explicitly, the reduced phase space $\mathfrak{P}_{\vec{A}}$ is defined by
\begin{eqnarray}
\mathfrak{P}_{\vec{A}}&:=&\bigg\{(A_1V^{IJ}_1,A_2V^{IJ}_2,...,A_nV^{IJ}_n)\in\times_{\imath=1}^n Q_{D-1}(A_\imath)
| \sum_{\imath=1}^nA_\imath V^{IJ}_\imath=0\bigg\}/SO(D+1),
\end{eqnarray}
and the Poisson structure on $\mathfrak{P}_{\vec{A}}$ is simply obtained via the symplectic reduction of the Poisson structure on $\times_{\imath=1}^n Q_{D-1}(A_\imath)$. Geometric quantization can be applied again to obtain the new Hilbert space $\mathcal{H}^{'c.}_{\vec{N}}$ of the $SO(D+1)$ invariant configurations of the faces satisfying the closure conditions. Alternatively, we may instead impose the corresponding quantum Gaussian constraints on the quantum space of the faces $\times_{\imath=1}^n\mathfrak{H}^{D+1,N_\imath}$. Recall that we have found the solution space $\mathcal{H}^{c.}_{\vec{N}}$ for these quantum constraints in the space $\times_{\imath=1}^n\mathfrak{H}^{D+1,N_\imath}$ as the images of the group-averaging map expressed in \eqref{group ave}. As it turns out, the Guillemin-Sternbergs theorem \cite{Guillemin} for the symplectic reduction implies that, the two operations-- the quantization and the imposition of the Gaussian (or Closure) constraints-- actually commute upon the original phase space $\times_{\imath=1}^n Q_{D-1}(A_\imath)$. Therefore, the obtained Hilbert spaces are identical as $\mathcal{H}^{'c.}_{\vec{N}}=\mathcal{H}^{c.}_{\vec{N}}$. Thus the exact correspondence holds between the impositions of the Gaussian constraint at the classical and quantum levels, as pictured in the following figure by the steps $(1)$ and $(2)$.

The second stage of imposing the simplicity constraints can no longer be treated in the same manner. The imposition of the classical simplicity constraints on the phase space $\mathfrak{P}_{\vec{A}}$, pictured as the step $(3)$ in the figure, leads to the following space. Consider n bi-vectors $\{A_1V^{IJ}_1,A_2V^{IJ}_2,...$ $...,A_nV^{IJ}_n\}$,
%in the tangent vector space (as a sub-space of $so(D+1)$) of arbitrary point $p_{\mathcal{N}^I}\in S^D$,
in the case of $V_\imath^{IJ}=\sqrt{2}\mathcal{N}^{[I}V_\imath^{J]}$ and $ \sum_{\imath=1}^nA_\imath V^{IJ}_\imath=0$, which thus corresponds to a D-polytope $p(\vec{A},\vec{V}_{IJ})$ with areas $\vec{A}=\{A_1,A_2,...,A_n\}$ and norms $\vec{V}_{IJ}=\{V^{IJ}_1,V^{IJ}_2,...,V^{IJ}_n\}$ for the (D-1)-faces. For the (D-1)-faces with arbitrary normal bi-vectors $V_\imath^{IJ}$, the space of shapes of a D-polytope with a given set of areas $\{A_1,A_2,...,A_n\}$ is given by
\begin{equation}
\mathfrak{P}^{s.}_{\vec{A}}=\bigg\{(A_1V^{IJ}_1,A_2V^{IJ}_2,...,A_nV^{IJ}_n)\in\times_{\imath=1}^n Q_{D-1}(A_\imath)
| \sum_{\imath=1}^nA_\imath V^{IJ}_\imath=0, V_\imath^{[IJ}V_\jmath^{KL]}=0\bigg\}/SO(D+1).
\end{equation}
The space of polytopes $\mathfrak{P}^{s.}_{\vec{A}}$, though well-defined, is not a phase space, and to our knowledge there has not been a valid approach of the direct quantization of this space. On the other side at the quantum level, we look into the imposition of the vertex quantum simplicity constraints on the Hilbert space $\mathcal{H}^{c.}_{\vec{N}}$, which is pictured as the step $(4)$ in the figure. For this step we compare two kinds of solutions of the vertex quantum simplicity constraints. One is the B-C intertwiners $\mathcal{I}^{B.C.}_{\vec{N}}$ as the strong solutions, and the other is our weak solutions of the coherent intertwiner space $\mathcal{H}^{s.c.}_{\vec{N}}$. Since the direct quantization of the space $\mathfrak{P}^{s.}_{\vec{A}}$ is out of reach, in this context the quantum Hilbert space of the solutions for which we are looking is only guided by the principle of the optimal correspondence. It would become clear that it is our weak solution space $\mathcal{H}^{s.c.}_{\vec{N}}$ that has the desired correspondence with the space $\mathfrak{P}^{s.}_{\vec{A}}$.

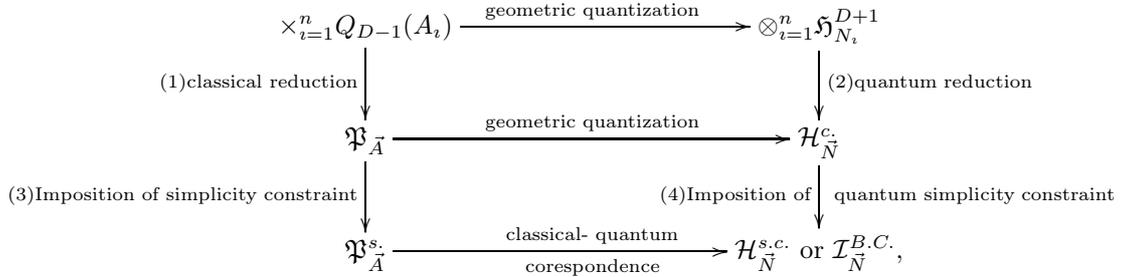
$$\xymatrix{}
\xymatrix@C=3.5cm{
\times_{\imath=1}^nQ_{D-1}(A_\imath)
\ar[d]_{\textrm{(1)classical\ reduction}} \ar[r]^{\textrm{geometric\ quantization}} &\otimes_{\imath=1}^n\mathfrak{H}^{D+1}_{N_\imath} \ar[d]^{\textrm{(2)quantum\ reduction}}\\
\mathfrak{P}_{\vec{A}}\ar[d]_{\textrm{(3)Imposition\ of\ simplicity\ constraint}} \ar[r]^{\textrm{geometric\ quantization}}& \mathcal{H}^{c.}_{\vec{N}} \ar[d]_{\textrm{(4)Imposition\ of}}^{\textrm{\ quantum\ simplicity\ constraint}} \\
\mathfrak{P}^{s.}_{\vec{A}} \ar[r]^{\textrm{classical- quantum}}_{\textrm{corespondence}}& \mathcal{H}^{s.c.}_{\vec{N}}\ \textrm{or}\ \mathcal{I}^{B.C.}_{\vec{N}} ,}$$

Note that an arbitrary classical state in $Q_{D-1}(A_1)$ can emerge from a coherent state in $\mathfrak{H}^{D+1}_{N_1}$ in the classical limit.
Here we also hope that an arbitrary classical state in $\mathfrak{P}^{s.}_{\vec{A}}$ can be realized based on a coherent state
%in the quantum reduction space of $\mathfrak{H}^{D+1}_{N_1} \times...\times\mathfrak{H}^{D+1}_{N_n}$ in the same limit,
in a final solution quantum space in a proper sense, which is a subspace of the quantum reduction space of $\mathfrak{H}^{D+1}_{N_1} \times...\times\mathfrak{H}^{D+1}_{N_n}$. Obviously, the strong solution space $\mathcal{I}^{B.C.}_{\vec{N}}$ would not provide such correspondence since it is only one dimensional. In contrast, our space $\mathcal{H}^{s.c.}_{\vec{N}}$ may be a reasonable ``quantum space" (or ``quantum counterpart" as we discussed at the beginning of this section) of $\mathfrak{P}^{s.}_{\vec{A}}$ in the following meaning.
The equivalent class (up to $SO(D+1)$ rotation) of polytopes $p(\vec{A},\vec{V}_{IJ})$ (where $V^\imath_{IJ}=\sqrt{2}\mathcal{N}_{[I}V^\imath_{J]}$) in D-dimensional Euclidean space can be given as a classical limit of simple coherent intertwiner $\mathcal{I}^{s.c.}:=||\vec{N},\vec{V}_{IJ}\rangle=\int_{SO(D+1)}dg \bigotimes_{\imath=1}^ng|N_{\imath},V^{\imath}_{IJ}\rangle$ with $\sqrt{N_\imath(N_\imath+D-1)}=A_\imath$ (up to a constant), with $\mathfrak{H}^{D+1}_{N_\imath}\ni|N_{\imath},V^{\imath}_{IJ}\rangle$ being the exact quantization of $Q_{D-1}(A_\imath)\ni A_\imath V_\imath^{IJ}$. A different point of view can be seen from the correspondence between the classical quantities of the faces and their operator representation in $\otimes_{\imath=1}^n\mathfrak{H}^{D+1}_{N_\imath}$. In the previous stage, the generators $X_\imath^{IJ}$ of $SO(D+1)$ acting on each simple representation space $\mathfrak{H}^{D+1}_{N_\imath}$ has been understood as the quantization of the vectors $A_{\imath}V_{\imath}^{IJ}$ up to a constant factor. In LQG the dimensionful constant is set to be $8\sqrt{2}\pi\beta(l_p^{(D+1)})^{D-1}$, where $\beta$ is the Immirzi parameter and $l_p^{(D+1)}$ is the Planck length in $(1+D)$-dimensional space-time. This gives
\begin{equation}
A_{\imath}V_{\imath}^{IJ} \quad\longmapsto\quad\frac{1}{\sqrt{2}}\beta\kappa\hbar\mathbf{i}X^{IJ}_\imath=8\sqrt{2}\pi\beta(l_p^{(D+1)})^{D-1}\mathbf{i}X^{IJ}_\imath,
\end{equation}
and accordingly the closure condition and simplicity constraint are promoted to operator equations as
\begin{eqnarray}
\sum_{\imath=1}^nA_\imath V^{IJ}_\imath=0 &\longmapsto&\sum_{\imath=1}^nX^{IJ}_\imath=0,\\\nonumber
V_\imath^{[IJ}V_\jmath^{KL]}=0&\longmapsto&X_\imath^{[IJ}X_\jmath^{KL]}=0.
\end{eqnarray}
The resulted quantum conditions are indeed the defining conditions of our space, which contain \eqref{zero quadratic expt}.

This choice of solution of simplicity constraint and the meaning of quantum counterpart can be understood by the toy model in quantum mechanics, which was introduced at the beginning of section 3. We considered a particle moving on a unit 2-sphere. The space of its angular momentum $\vec{J}$ with norm $j$ is given by 2-sphere ${}^2S_j$ with radius $j$ equipped with its invariant volume 2-form. This phase space ${}^2S_j$ could be geometrically quantized as the representation space $V^{(j)}$ with $j=0,1/2,1,3/2,...$ of $SU(2)$. Classically, we can impose the condition $J_z=0$ on ${}^2S_j$ and get the subspace ${}^1S^{\bar{z}}_j\in{}^2S_j$, which is the circle composed by the point with $J_z=0$ in ${}^2S_j$. Analogously, we can impose the quantized formulation $\hat{J}_z=0$ on $V^{(j)}$ and get the state $|j,0\rangle$ as the only strong solution of $\hat{J}_z=0$. It is easy to see that the strong solution space of the condition $\hat{J}_z=0$ does not have enough degrees of freedom to describe the classical state in ${}^1S^{\bar{z}}_j$. Here we have another choice given by ${}^1S^{\bar{z}}_{c.j}$ which is composed of all the coherent states like $|j,\vec{n}_{\bar{z}}\rangle$, where $\vec{n}_{\bar{z}}$ represents any unit vector orthogonal to $J_z$. It is easy to see that $\langle j,\vec{n}_{\bar{z}}|\hat{J}|j,\vec{n}_{\bar{z}}\rangle=j\vec{n}_{\bar{z}}\in{}^1S^{\bar{z}}_{j}$ and $|j,\vec{n}_{\bar{z}}\rangle$ minimal the uncertainty of $\hat{J}$ in $V^{(j)}$. Thus the coherent state $|j,\vec{n}_{\bar{z}}\rangle$ peaks $\vec{J}$ at the point $j\vec{n}_{\bar{z}}\in{}^1S^{\bar{z}}_{j}$. Now it is obvious that all the classical states in ${}^1S^{\bar{z}}_{j}$ could be realized if we neglect the quantum uncertainty. Hence ${}^1S^{\bar{z}}_{c.j}$ is a more reasonable ``quantum space'' (or the ``quantum counterpart'' as we used above) corresponding to classical state space ${}^1S^{\bar{z}}_{j}$. These arguments can be summarized in the following figure.

$$\xymatrix{}
\xymatrix@C=3.5cm{
{}^2S_j
\ar[d]^{J_z=0} \ar[r]^{\textrm{geometric\ quantization}} &V^{(j)} \ar[d]^{\hat{J}_z=0}\\
{}^1S^{\bar{z}}_{j} \ar[r]^{\textrm{classical\ quantum}}_{\textrm{correspondence}}& {}^1S^{\bar{z}}_{c.j}. }$$

\section{Comparison of two kinds of coherent intertwiner spaces for (1+3)-dimensional LQG}

While the universal connection formulation of $(D+1)$-dimensional GR has to use the $SO(D+1)$ gauge group with the additional simplicity constraints, the $(1+3)$-dimensional case enjoys a simpler formulation using directly the $SU(2)$ gauge group without the simplicity constraints. Since the latter special formulation is the prevailing one for $(1+3)$-dimensional LQG, it is important to compare our solution space with the well-known $SU(2)$ coherent intertwiner solution space of the $SU(2)$ quantum Gaussian constraints. Particularly, we want to check if the two agree in the quantum geometric information encoded in each vertex of the respective spin-network states. In the universal formulation with $D=3$ the gauge group is $SO(4)$, or equivalently its double covering $Spin(4)$, with which the loop quantization has been studied in Bodendorfer-Thiemann approach of LQG.

First we start from a natural correspondence between $\mathcal{H}^{s.c.}_{Spin(4)}$ and $\mathcal{H}^{s.c.}_{SU(2)}$:
\begin{equation}\label{coresp}
||\vec{j},\vec{\vec{n}}\rangle\Leftrightarrow p(\vec{j},\vec{\vec{n}})\longleftrightarrow p(\vec{N},\vec{V}_{IJ})\Leftrightarrow||\vec{N},\vec{V}_{IJ}\rangle, \quad \vec{j}=\frac{1}{2}\vec{N},\quad V_{IJ}^\imath=V_{IJ}(\vec{n}^\imath),
\end{equation}
where $||\vec{j},\vec{\vec{n}}\rangle:=\int_{SU(2)}g|j_1,\vec{n}_1\rangle\otimes...\otimes g|j_n,\vec{n}_n\rangle dg$ is the $SU(2)$ coherent intertwiner corresponding to the equivalent class of the polyhedrons $p(\vec{j},\vec{\vec{n}})$ that have 2-faces with areas $(j_1,...,j_n)$ and unit normal vectors $(\vec{n}_1,...,\vec{n}_n)$, the factor $\frac{1}{2}$ is chosen to make sure the corresponding 2-faces' area spectra given by the two kinds of intertwiners agree (up to a constant), and the map $V_{IJ}(\cdot): su(2)\mapsto so(4)$ satisfies
\begin{eqnarray}
<\vec{n}^\imath,\vec{n}^\jmath>=<V_{IJ}(\vec{n}^\imath), V_{IJ}(\vec{n}^\jmath)>,\quad V_{[IJ}(\vec{n}^\imath) V_{KL]}(\vec{n}^\jmath)=0,\quad \forall \imath,\jmath.
\end{eqnarray}
Here the inner product $<\cdot,\cdot>$ is given by the Cartan-Killing metrics of $su(2)$ and $so(4)$ respectively. To understand the meaning of the above correspondence, we note that the space of shapes of 3-polyhedron could be given by either of the following two ways:
\begin{eqnarray}
\mathfrak{P}^{s.}_{\vec{N}}&:=&\{(N_1V^{IJ}_1,N_2V^{IJ}_2,...,N_nV^{IJ}_n)\in Q(N_1)\times Q(N_2) \times
\\\nonumber&&...\times Q(N_n)
| \sum_{\imath=1}^nN_\imath V^{IJ}_\imath=0,\quad V_\imath^{[IJ}V_\jmath^{KL]}=0\}/SO(4)\\\nonumber
\mathcal{S}_{\vec{j}}&:=&\{(j_1\vec{n}_1,...,j_n\vec{n}_n)|\sum_{\imath=1} ^nj_\imath\vec{n}_\imath=0\}/SU(2),
\end{eqnarray}
which correspond to the $SO(4)$ formulation and the $SU(2)$ formulation respectively. These two spaces $\mathfrak{P}^{s.}_{\vec{N}}$ and $\mathcal{S}_{\vec{j}}$ have the same dimension of $(2n-6)$. It is easy to see that there exists a one to one and onto map between these two spaces, such that an element in $\mathfrak{P}_{\vec{N}}$ is mapped to the element in $\mathcal{S}_{\vec{j}}$ with the same shape by $\vec{j}=c\vec{N}$, where $c>0$ is an arbitrary real number.

After establishing this natural correspondence between the coherent states in the two intertwiner spaces, we now compare their corresponding inner products at the quantum level. Since the $SO(4)$ group has the double covering group given by $Spin(4)=SU(2)\times SU(2)$, the $SO(4)$ simple coherent intertwiner space can be equivalently given by $\mathcal{H}^{s.c.}_{Spin(4)}$. We notice the fact that $so(4)\cong su(2)_{\text{L}}\oplus su(2)_{\text{R}}$ and correspondingly $|N_\imath,V^{IJ}_\imath\rangle=|j^\imath_L,\vec{n}^\imath_L\rangle \otimes|j^\imath_R,\vec{n}^\imath_R\rangle$ with $j^\imath_L=j^\imath_R=N/2$, $\vec{n}^\imath_L=\frac{1}{\sqrt{2}}(V^{IJ}_\imath+\bar{V}^{IJ}_\imath)$ and $\vec{n}^\imath_R=\frac{1}{\sqrt{2}}(V^{IJ}_\imath-\bar{V}^{IJ}_\imath)$, where $\bar{V}^{IJ}_\imath:=\frac{1}{2}\epsilon^{IJKL}V_{KL}^\imath$ commute with $V^{IJ}_\imath$ in $so(4)$. Hence $|N_\imath,V^{IJ}_\imath\rangle$ is an eigenstate of the projection of the $su(2)_{\text{L}}$-valued vector operator $\vec{\tau}^\imath_{\text{L}}:=\frac{X_\imath^{IJ}+1/2\epsilon^{IJ}_{\ \ KL}X_\imath^{KL}}{2}$ on $\vec{n}^\imath_L$ with eigenvalue proportional to $\mathbf{i}j^\imath_L$, and it is also an eigenstate of the projection of the $su(2)_{\text{R}}$-valued vector operator $\vec{\tau}^\imath_{\text{R}}:=\frac{X_\imath^{IJ}-1/2\epsilon^{IJ}_{\ \ KL}X_\imath^{KL}}{2}$ on $\vec{n}^\imath_R$ with eigenvalue proportional to $\mathbf{i}j^\imath_R$. Importantly, the satisfied simplicity condition $V^{IJ}_\imath V^{KL}_\jmath\epsilon_{IJKL}=0$ implies that we have$<\vec{n}_L^\imath,\vec{n}_L^\jmath> =<\vec{n}_R^\imath,\vec{n}_R^\jmath> =<V^{IJ}_\imath,V^{IJ}_\jmath>$, and thus the decomposition here is through the natural correspondence we have introduced above. The inner product in $\mathcal{H}_{\vec{N}}^{s.c.}$ is directly given by the inner products among the constituent states $||\vec{N},\vec{V}^{IJ}\rangle$, which can be evaluated in the corresponding form as
\begin{eqnarray}
&&\langle\vec{N},\vec{V}^{IJ}||\vec{N},\vec{V}'^{IJ}\rangle \\\nonumber &:=& \int_{Spin(4)}\int_{Spin(4)}dgdh\prod_{\imath=1}^n\langle N_\imath,V^{IJ}_\imath|h^{-1}g|N_\imath,V'^{IJ}_\imath\rangle\\\nonumber &=&\int_{Spin(4)}\int_{Spin(4)}dgdh\prod_{\imath=1}^n\langle j^\imath_L,\vec{n}^\imath_L|h^{-1}g|j^\imath_L,\vec{n}'^\imath_L\rangle \langle j^\imath_R,\vec{n}^\imath_R|h^{-1}g|j^\imath_R,\vec{n}'^\imath_R\rangle\\\nonumber &=&\int_{SU(2)_L}\int_{SU(2)_R}dg_Ldg_R\prod_{\imath=1}^n\langle j^\imath_L,\vec{n}^\imath_L|g_L|j^\imath_L,\vec{n}'^\imath_L\rangle \langle j^\imath_R,\vec{n}^\imath_R|g_R|j^\imath_R,\vec{n}'^\imath_R\rangle
\end{eqnarray}
with the label choices satisfying $V^{[IJ}_\imath V'^{KL]}_\imath=0$ and the identity $<\vec{n}^\imath_L,\vec{n}'^\imath_L>=<\vec{n}^\imath_R,\vec{n}'^\imath_R>=<V^{IJ}_\imath, V'^{IJ}_\imath>=\cos\theta_\imath$, while the inner product of $||\vec{j},\vec{\vec{n}}\rangle$ in $\mathcal{H}_{\vec{j}}^{c.}$ is given by
\begin{eqnarray}
\langle\vec{j},\vec{\vec{n}}||\vec{j},\vec{\vec{n}}'\rangle&:=&\int_{SU(2)}\int_{SU(2)}dgdh\prod_{\imath=1}^n\langle j^\imath,\vec{n}^\imath|h^{-1}g|j^\imath,\vec{n}'^\imath\rangle \\\nonumber
&=&\int_{SU(2)}dg\prod_{\imath=1}^n\langle j^\imath,\vec{n}^\imath|g|j^\imath,\vec{n}'^\imath\rangle.
\end{eqnarray}
Therefore, using $j_\imath=\frac{1}{2}N_\imath$, one gets
\begin{eqnarray}
(\langle\vec{j},\vec{\vec{n}}||\vec{j},\vec{\vec{n}}'\rangle)^2 &=&\langle\vec{N},\vec{V}^{IJ}||\vec{N},\vec{V}'^{IJ}\rangle.
\end{eqnarray}
We thus conclude that, establishing a one-to-one correspondence at the classical level of shapes of polytopes, the map between the two intertwiner spaces $\mathcal{H}^{c.s.}_{Spin(4)}$ and $\mathcal{H}^{c.}_{SU(2)}$ based on the correspondence relation (\ref{coresp}) is not unitary, and thus not an identification at the quantum level.

Finally, we briefly comment on the relation of our solution space to the strong solution space of the maximal non-anomalous set of the quantum simplicity constraints, which was promoted by Bodendofer and Thiemann in \cite{Bodendorfer:2011onthe}. The standard orthonormal basis for the space of gauge invariant intertwiners is given by the set of generalized Clebsch-Gordon invariant tensors. As it is well-known, with a given set of representations assigned to the edges connected to a vertex in a spin-network state, the set of valid invariants tensors for this vertex is given by the set of distinct ways to re-couple those representations into one trivial representation. Moreover, each specific way of the re-coupling may be encoded by a specific network of the ``internal edges'' inside of the vertex, thus these basis elements can be denoted as
$\{i_{\vec{j}}^{(k_1,k_2,...)},...,i_{\vec{j}}^{(k'_1,k'_2,...)}\}$, each of which related to a given re-coupling scheme, where $(k_1,k_2,...), (k'_1,k'_2,...)$ are the labelling of the inner edges of this given re-coupling scheme which satisfy the rules of Clebsch-Gordan decomposition. For the $SU(2)$ formulation, an arbitrary element in $\mathcal{H}_{\vec{j}}^{c.}$ can be decomposed with this orthogonal basis as
\begin{equation}
||\vec{j},\vec{\vec{n}}\rangle=\sum_{k_1,k_2,...}c^{\vec{j},\vec{\vec{n}} }_{k_1,k_2,...}i_{\vec{j}}^{(k_1,k_2,...)}.
\end{equation}
Similarly, in the $SO(4)$ case we have the decomposition
\begin{equation}
||\vec{N},\vec{V}^{IJ}\rangle= ||\vec{j}_L,\vec{\vec{n}}_L\rangle \otimes ||\vec{j}_R,\vec{\vec{n}}_R\rangle=(\sum_{k_1,k_2,...}c^{\vec{j},\vec{\vec{n}} }_{k_1,k_2,...}i_{\vec{j}}^{(k_1,k_2,...)})_L\otimes (\sum_{k_1,k_2,...}c^{\vec{j},\vec{\vec{n}} }_{k_1,k_2,...}i_{\vec{j}}^{(k_1,k_2,...)})_R,
\end{equation}
where $\vec{j}=\vec{j}_L=\vec{j}_R$, and $\frac{<\vec{n}_L^\imath,\vec{n}_L^\jmath>}{|\vec{n}_L^\imath|\cdot|\vec{n}_L^\jmath|} =\frac{<\vec{n}_R^\imath,\vec{n}_R^\jmath>}{|\vec{n}_R^\imath|\cdot|\vec{n}_R^\jmath|} =\frac{<\vec{n}^\imath,\vec{n}^\jmath>}{|\vec{n}^\imath|\cdot|\vec{n}^\jmath|}$.
This decomposition provides us an expression of our simple coherent intertwiner solutions in terms of the recoupling operations. And also this form can be compared with the solutions of maximal commuting subset of vertex simplicity constraints, which can be span by such kind of states
\begin{equation}
\mathcal{I}^{\textrm{m.c.}}_{{\vec{j}},{(k_1,k_2,...)}}=(i_{\vec{j}}^{(k_1,k_2,...)})_L\otimes ( i_{\vec{j}}^{(k_1,k_2,...)})_R
\end{equation}
in $\mathcal{H}_{\vec{j},L}\otimes \mathcal{H}_{\vec{j},R}$ for the given re-coupling scheme.
Now it is obvious that both the simple coherent intertwiner solution $||\vec{N},\vec{V}^{IJ}\rangle$ and the solution $\mathcal{I}^{\textrm{m.c.}}_{{\vec{j}},{(k_1,k_2,...)}}$ of maximal commutate subset of vertex simplicity constraints can be regarded as the direct product of two same $SU(2)$ intertwiners, while the difference is that $||\vec{N},\vec{V}^{IJ}\rangle$ is the direct product of two same $SU(2)$ coherent intertwiners $||\vec{j},\vec{\vec{n}}\rangle$ but $\mathcal{I}^{\textrm{m.c.}}_{{\vec{j}},{(k_1,k_2,...)}}$ is the direct product of two same $SU(2)$ intertwiner bases with a given re-coupling scheme. Hence the two kinds of solutions locate in different subspaces of $\mathcal{H}_{\vec{j},L}\otimes \mathcal{H}_{\vec{j},R}$ in general.

\section{Conclusion and Remark}

%It is worth to have a general discussion of simplicity constraints to make sure we can understand it more clearly. In classical theory, the gauge transformation orbit of simplicity constraint lie in constraint surface of Gauss constraints and simplicity constraints as we have a first class constraint system, we can check that the $\pi^{aIJ}$ field is invariant and only the project components $\bar{A}_{bKL}$ of $A_{bKL}$ field which is given by $\bar{A}_{bIJ}:=\bar{\eta}^K_I[n^M]\bar{\eta}^L_J[n^M]A_{bKL}$, $\bar{\eta}^K_I[n^M]:=\delta^K_I-n^Kn_I$, $n_I=\mathcal{N}^I[\pi^{aIJ}]$ would change along the gauge transformation orbit of simplicity constraint. But the case is totally different in quantum theory as the algebra of quantum quadratic simplicity constraints are not closed any more, which means the

For the universal formulation of canonical LQG, we have introduced a new approach to analyze the anomalous standard quantum simplicity constraints, using our weak solution space to the constraints with the degrees of freedom having clear geometric interpretations. In the space of cylindrical functions, we have identified a specific subspace spanned by a set of states with the simple representations assigned to the edges, solving the edge simplicity constraints, and with the vertex intertwiner space $\check{\mathcal{H}}_{\vec{N}_v}^{s.c.\delta}$ or $\check{\mathcal{H}}_{\vec{N}_v}^{s.c.}$ weakly solving both the vertex quantum simplicity constraints and the quantum Gaussian constraints. The vertex intertwiner space is given by the coherent intertwiners peaking at the bi-vector values satisfying the classical vertex simplicity conditions, thereby it has the vanishing expectation values for the vertex simplicity constraints. Then, we showed how to impose the quantum Gaussian constraints strongly by applying the rigging map and find the $SO(D+1)$ gauge invariant vertex intertwiner space $\mathcal{H}_{\vec{N}_v}^{s.c.}$, assumed to be faithfully represented by $\check{\mathcal{H}}_{\vec{N}_v}^{s.c.}$. Remarkably, the degrees of freedom of this gauge-invariant intertwiner space has a natural correspondence with the space of shape of the Euclidean D-polytopes, which indeed can be viewed as the building blocks for the Riemannian spatial geometry arising from imposing the Gaussian and simplicity constraints in the phase space of the $SO(D+1)$ gauge theory.

Our weak solutions are constructed from the $SO(D+1)$ coherent states with minimal uncertainties in the flux operators $X_{IJ}$. This method is applicable to both of the quadratic and the linear quantum simplicity constraints at the $SO(D+1)$ gauge-fixed level, such that the expectation values of both versions of the constraints are exactly zero in the space $\check{\mathcal{H}}_{s.c.}^{v,\delta}$. For the same underlying reason, the expectation values of the anomalous commutators of the quantum constraints are also zero. At the gauge invariant level, the expectation values for the quadratic constraints only tend to zero in the large $N$ limit. We have shown that, the most important point for these to happen is that the expectation values of the building factors $X_{IJ}$ of the quantum simplicity constraint operators in the $SO(D+1)$ coherent states have minimized quantum uncertainty which could be ignored in large $N$ limit.

In contrast to the strong solution space of the quantum simplicity constraints, which lacks the physical degrees of freedom, the degrees of freedom in our weak solution space are labelled by the shapes of the classical polytopes dual to the vertices of the spin-network states. Thus this space  can be thought of as a space of quantum polytopes. We also noted that our solution space cannot be obtained from a direct quantization of the classical phase space of polytopes, but rather it is the ``quantum counterpart" of the space of shapes of polytopes with a fixed set of areas for the $(D-1)$-faces, such that any classical polytope can be given by a simple coherent intertwiner in a classical limit. In this sense, our weak solution states may be more suitable to describe quantum geometry compared to those using the B-C intertwiners.

For the special $D=3$ case allowing the $SU(2)$ formulation, we compared our $Spin(4)$ simple coherent intertwiners with $SU(2)$ coherent intertwiners, both giving the quantum polyhedrons in 3-dimensional Euclidean space. A correspondence between these two kinds of intertwiners is established by matching their shapes and their area spectra (up to a constant) of the 2-surfaces. However, the correspondence does not give a unitary map between the two intertwiner spaces. This implies that these two quantum theories might be distinct at the quantum kinematic level.

Our analysis of the simple coherent intertwiners also points to a few open issues for future study. First, beyond the qualitative argument given in the Appendix, we should look for a precise proof for the statement that the expectation values of the quantum quadratic simplicity constraints vanish in large $N$ limit for the gauge invariant simple coherent states. Second, while the simple coherent intertwiner solutions have good classical limit, we would also like to understand the quantum properties of the solutions; for instance, it is interesting to observe that the minimized relative uncertainty of flux operators inevitably becomes large when $N$ is small, indicating highly quantum solutions far from the constraint surface of classical simplicity constraints in phase space. Third, the difference in the quantum properties of the simple $Spin(4)$ coherent intertwiners and the $SU(2)$ coherent intertwiners are yet to be studied, for better understanding the relation between the two formulations of (1+3)-dimensional LQG. Lastly, since there is no guarantee that the anomalous simplicity constraints will commute with the Hamiltonian constraints in quantum theory, it remains to be investigated how much of the geometric information in the simple coherent intertwiners can survive at the level of the physical Hilbert space.

\section*{Acknowledgments}
We would like to thank Shupeng Song and Cong Zhang for discussions and useful comments
on a draft of this paper. This work is supported by the National Natural Science Foundation of China (NSFC) with Grants No. 11875006.

\appendix
\section{On the expectation value of quadratic simplicity constraint operator for gauge invariant simple coherent intertwiners}
%Though we claimed that the gauge invariant simple coherent intertwiner $\mathcal{I}_{v,\vec{N}_v,\vec{g}_D}^{s.c.}$ is weak solution of quadratic simplicity constraints, but the expected value of building blocks $R_{e_{\jmath_1}}^{[IJ}R_{e_{\jmath_2}}^{KL]} $ of quantum quadratic simplicity constraint do not vanish when we impose them weakly as what we do in section 4.
In contrast to  the gauge-fixed simple coherent intertwiners $\mathcal{\check{I}}_{v,\vec{N}_v,\vec{g}_D}^{s.c.}$, for gauge invariant simple coherent intertwiners the expectation value of $R_{e_{\jmath_1}}^{[IJ}R_{e_{\jmath_2}}^{KL]} $ do not vanish in general. To discuss the weak imposition of quantum quadratic simplicity on gange invariant simple coherent intertwiners, we first consider the property of $SO(D+1)$ coherent states. An orthogonal coordinate system $(x_1,x_2,...,x_{D+1})$ in $(D+1)$-dimensional Euclidean space and a spherical coordinate system $\vec{\xi}=(\xi_1,\xi_2,...,\xi_{D})$ (wherein $0\leq \xi_1<2\pi, 0\leq\xi_2,\xi_3,...,\xi_{D}\leq \pi$) are related by
\begin{eqnarray}
x_{D+1}&=&r\cos\xi_D,\\\nonumber
x_D&=&r\sin\xi_D\cos\xi_{D-1},\\\nonumber
x_{D-1}&=&r\sin\xi_D\sin\xi_{D-1}\cos\xi_{D-2},\\\nonumber
&&...\\\nonumber
x_2&=&r\sin\xi_D\sin\xi_{D-1}...\sin\xi_{2}\sin\xi_1,\\\nonumber
x_1&=&r\sin\xi_D\sin\xi_{D-1}...\sin\xi_{2}\cos\xi_1.
\end{eqnarray}
Then the coherent state $|N\textbf{e}_1\rangle$ of $SO(D+1)$ could be given as the homogenous harmonic function,
\begin{equation}
\Xi^{N\textbf{e}_1}_{D+1}(\vec{\xi}):=C_N\sin^N\xi_D\sin^N\xi_{D-1} ...\sin^N\xi_{2}e^{\textbf{i}N\xi_1},
\end{equation}
where $C_N$ is the normalization constant. Notice that the function $\sin^N\xi$ is sharply peaked at $\xi=\frac{\pi}{2}$ in large $N$ limit. Hence for $N\rightarrow \infty$, $\Xi^{N,\textbf{e}_1}_{D+1}(\vec{\xi})$ is peaked at a circle ${}^1\!S_{1,2}$ in $S^D$, which is labelled by $\xi_2=\xi_{3}=...=\xi_{D}=\frac{\pi}{2}$. Also, this circle is the intersection of $S^D$ and the 2-plane which contains the original point and is parallel with $\frac{\partial}{\partial x_1}, \frac{\partial}{\partial x_2}$. This discussion is valid for an arbitrary coherent state $|N,g\rangle$. For $N\rightarrow \infty$, $|N,g\rangle$ can be regarded as a harmonic homogenous function on $S^D$ and it is peaked at the circle ${}^1\!S(g)$ which is the intersection of $S^D$ and the 2-plane which contains the original point and is parallel with $g\frac{\partial}{\partial x_1}, g\frac{\partial}{\partial x_2}$. It follows qualitatively that the product $\langle N,g'|N,g\rangle$ tend to zero for $N\rightarrow \infty$, if the circles ${}^1\!S(g)$ and ${}^1\!S(g')$ are not identical. In fact, there are two situations that circles ${}^1\!S(g)$ and ${}^1\!S(g')$ are not identical: (1). ${}^1\!S(g)$ and ${}^1\!S(g')$ has no intersection point. Based on the above property of the coherent states, we can always find $N$ that is large enough to separate the two wave functions $|N,g\rangle$ and $|N,g'\rangle$ on $S^D$. Then the product $\langle N,g'|N,g\rangle$ will tend to zero in large $N$ limit. (2). ${}^1\!S(g)$ and ${}^1\!S(g')$ intersect with each other and can be put into a 2-sphere. We can adjust the coordinates $(x_1,x_2,...,x_{D+1})$ so that this 2-sphere is coordinated by $(x_1,x_2,x_3)$, and $|N,g\rangle$ and $|N,g'\rangle$ are given by following two homogeneous harmonic functions on $S^D$,
 \begin{equation}\label{rel1}
\Xi^{N,\textbf{e}_1}_{D+1}(\vec{\xi})=\frac{C_N}{c_N}\sin^N\xi_D\sin^N\xi_{D-1} ...\sin^N\xi_{3}\Xi^{N,N}_{3}(\xi_{2},\xi_{1}),
\end{equation}
and
\begin{equation}\label{rel2}
\Xi^{N,\bar{g}}_{D+1}(\vec{\xi})=\frac{C_N}{c_N}\sin^N\xi_D\sin^N\xi_{D-1} ...\sin^N\xi_{3}\Xi^{N,\bar{g}}_{3}(\xi_{2},\xi_{1}),
\end{equation}
where $\Xi^{N,N}_{3}(\xi_{2},\xi_{1})$ and $\Xi^{N,\bar{g}}_{3}(\xi_{2},\xi_{1})$ are $SO(3)$ coherent states which are given by the harmonic functions on the 2-sphere, with $c_N$ being their normalization constant, and $\bar{g}\in SO(3)$ rotates $\Xi^{N,N}_{3}(\xi_{2},\xi_{1})$ to $\Xi^{N,\bar{g}}_{3}(\xi_{2},\xi_{1})$. The two $SO(3)$ coherent states can also be expressed as $|N,\frac{\partial}{\partial x_3}\rangle$ and $|N, \vec{\bar{n}}\rangle$ with $\vec{\bar{n}}=\bar{g}\frac{\partial}{\partial x_3}$. Let the coherent states be normalized as
\begin{equation}
\int_{S^D}d\mu(\xi_{D+1},...,\xi_3)d\mu(\xi_2,\xi_1)\Xi^{N,\bar{g}}_{D+1}(\vec{\xi})\overline{\Xi^{N,\bar{g}}_{D+1}(\vec{\xi})}=1,
\end{equation}
and
\begin{equation}
\int_{{}^2\!S}d\mu(\xi_2,\xi_1)\Xi^{N,\bar{g}}_{3}(\xi_2,\xi_1)\overline{\Xi^{N,\bar{g}}_{3}(\xi_2,\xi_1})=1,
\end{equation}
where $d\mu(\xi_{D+1},...,\xi_3)d\mu(\xi_2,\xi_1)$ is the normalization measure on $S^D$ and $d\mu(\xi_2,\xi_1)$ is the normalization measure on ${}^2\!S$. It then follows that
\begin{equation}\label{int}
\int_{S^D/{}^2\!S}d\mu(\xi_{D+1},...,\xi_3) |\frac{C_N}{c_N}\sin^N\xi_D\sin^N\xi_{D-1} ...\sin^N\xi_{3}|^2=1.
\end{equation}
 The $SO(3)$ coherent intertwiner $|j,\vec{n}\rangle$ satisfies \cite{GeneralizedCoherentStates}
\begin{equation}
\langle j,\vec{n}|g(\theta)g(\phi)|j,\vec{n}\rangle=e^{\textbf{i}j\phi}(\frac{1+\cos\theta}{2})^j,
\end{equation}
where $g(\phi)$ is an element of $ SO(2)_{\vec{n}}$ which preserves $\vec{n}$, $g(\theta)\in SO(3)/SO(2)_{\vec{n}}$, and $<g(\theta)\vec{n},\vec{n}>=\cos\theta$. Notice that the $SO(3)$ coherent state $|j,\vec{n}\rangle$ with $ j=N$ and $\vec{n}=\frac{\partial}{\partial x_3}$ of $SO(3)$ can also be given as a homogenous harmonic function,
 \begin{equation}
\Xi^{j,\vec{n}}(\xi_2,\xi_1):=c_j\sin^j\xi_{2}e^{\textbf{i}j\xi_1},\quad j=N, \vec{n}=\frac{\partial}{\partial x_3}.
\end{equation}
Based on Eqs. (\ref{rel1}), (\ref{rel2}) and (\ref{int}), we can conclude that
\begin{equation}
\langle N,\textbf{e}_1|g(\theta)g(\xi_1)|N,\textbf{e}_1\rangle=e^{\textbf{i}N\xi_1} (\frac{1+\cos\theta}{2})^N,
\end{equation}
where $g(\xi_1)$ is an element of $ SO(2)\in SO(3)$ which preserves $\vec{n}$, and by using bi-vector labeling we have $2n_{1,2}^{IJ}n^{1,2}_{IJ}(\theta)=\cos\theta$, with $n_{1,2}^{IJ}:=\delta^{[I}_1\delta^{J]}_2$ being a bi-vector in $\mathbb{R}^{D+1}$ and $n_{1,2}^{IJ}(\theta)$ given by rotating $n_{1,2}^{IJ}$ with $g(\theta)$ in the adjoint representation. The above result can be extended to more general case as
\begin{equation}\label{NEEN}
\langle N,\textbf{e}_1|g(\theta)(g(\xi_1)\times h)|N,\textbf{e}_1\rangle=e^{\textbf{i}N\xi_1} (\frac{1+\cos\theta}{2})^N,
\end{equation}
where $g(\xi_1)$ is an element of $ SO(2)$ which gives the rotation of the two-dimensional vector space spanned by $(
\frac{\partial}{\partial x_1},\frac{\partial}{\partial x_2})$, $h$ is an element of $ SO(D-1)$ which preserves $
\frac{\partial}{\partial x_1}$ and $\frac{\partial}{\partial x_2}$,  and $g(\theta)\in SO(D+1)/(SO(2)\times SO(D-1))$ with $2n_{1,2}^{IJ}n^{1,2}_{IJ}(\theta)=\cos\theta$ and $\vec{n}_{1,2}^{IJ}(\theta)$ being given by rotating $\vec{n}_{1,2}^{IJ}$ with $g(\theta)$ in the adjoint representation. It is obvious that Eq.\eqref{NEEN} will tend to zero in large $N$ limit.
The above discussion indicates that the matrix element function $\langle N,\textbf{e}_1|g|N,\textbf{e}_1\rangle$ on $SO(D+1)$ is sharply peaked on the subgroup $SO(2)\times SO(D-1)\subset SO(D+1)$ while $N\rightarrow\infty$. Now consider the following equation
\begin{eqnarray}
&&\langle\gamma,\mathcal{I}_{v,\vec{N}_v,\vec{g}_D}^{s.c.},...|R_{e_{\jmath_1}}^{[IJ}R_{e_{\jmath_2}}^{KL]} |\gamma,\mathcal{I}_{v,\vec{N}_v,\vec{g}_D}^{s.c.},...\rangle\\\nonumber &\simeq&\int_{SO(D+1)}\int_{SO(D+1)} \prod_{\imath\neq\jmath_1,\jmath_2,\imath=1}^{n_v} \langle N_{\imath},g^{\imath}_D|g|N_{\imath},g^{\imath}_D\rangle\\\nonumber
&&\cdot \langle N_{\jmath_1},g^{\jmath_1}_D|h^{-1}X^{[IJ}hg|N_{\jmath_1},g^{\jmath_1}_D\rangle\cdot \langle N_{\jmath_2},g^{\jmath_2}_D|h^{-1} X^{KL]}hg|N_{\jmath_2},g^{\jmath_2}_D\rangle dgdh,
\end{eqnarray}
where $\langle N_{\imath},g^{\imath}_D|g|N_{\imath},g^{\imath}_D\rangle=\langle N_{\imath},\textbf{e}_1|{g^{\imath}_D}^{-1}gg^{\imath}_D|N_{\imath},\textbf{e}_1\rangle$ is sharply peaked on $g^{\imath}_D(SO(2)\times SO(D-1)){g^{\imath}_D}^{-1}\subset SO(D+1)$ when $N_{\imath}\rightarrow\infty$. If $\mathcal{I}_{v,\vec{N}_v,\vec{g}_D}^{s.c.}$ was able to give a D-polytope as described in section 4, the function $\prod_{\imath\neq\jmath_1,\jmath_2,\imath=1}^{n_v} \langle N_{\imath},g^{\imath}_D|g|N_{\imath},g^{\imath}_D\rangle$ would be sharply peaked on identity $\textrm{Id.}\in SO(D+1)$ when $N_{\imath}\rightarrow\infty$. Notice that the factor $\langle N_{\jmath_1},g^{\jmath_1}_D|h^{-1}X^{[IJ}hg|N_{\jmath_1},g^{\jmath_1}_D\rangle\cdot \langle N_{\jmath_2},g^{\jmath_2}_D|h^{-1} X^{KL]}hg|N_{\jmath_2},g^{\jmath_2}_D\rangle$  vanishes while $g=\textrm{Id.}$. Therefore $\langle\gamma,\mathcal{I}_{v,\vec{N}_v,\vec{g}_D}^{s.c.},...|R_{e_{\jmath_1}}^{[IJ}R_{e_{\jmath_2}}^{KL]} |\gamma,\mathcal{I}_{v,\vec{N}_v,\vec{g}_D}^{s.c.},...\rangle$ would tend to zero when $N_{\imath}\rightarrow\infty$.

\bibliographystyle{unsrt}

\bibliography{coherentintertwiner16}

\end{document}